\documentclass[fleqn,usenatbib]{mnras}
\usepackage{newtxtext,newtxmath}
\usepackage[T1]{fontenc}
\usepackage{romannum}
\usepackage{graphicx} % Including figure files
\usepackage[tight,footnotesize]{subfigure}
\usepackage{amsmath}	% Advanced maths commands
\usepackage{multirow}
\usepackage{comment}
\usepackage{textcomp}
\usepackage[center]{caption}
\usepackage{hyperref}
\hypersetup{
    colorlinks=true,
    linkcolor=blue,
    filecolor=magenta,      
    urlcolor=blue,
    citecolor=blue,
}

\newcommand{\orb}{${\Omega}$\,} % Omega
\newcommand{\twoorb}{${2\Omega}$\,} % 2Omega
\newcommand{\s}{${\omega}$\,} % omega
\newcommand{\be}{$\omega$\,-\,$\Omega$\,} % omega - Omega
\newcommand{\twobe}{$2(\omega$\,-\,$\Omega)$\,} % omega - Omega
\newcommand{\threebe}{3($\omega$\,-\,$\Omega$\,)} % omega - Omega
\newcommand{\beplus}{$\omega$\,+\,$\Omega$\,} % omega + Omega
\newcommand{\sone}{$2\omega$\,-\,$\Omega$\,} % 2omega - Omega
\newcommand{\stwo}{$2\omega$\,-\,$3\Omega$\,} % 2omega - 3Omega
\newcommand{\newf}{$2\Omega$\,-\,$\omega$\,} % 2Omega - omega
\newcommand{\po}{P$_{\Omega}$\,} % P_Omega
\newcommand{\ptwoo}{P$_{2\Omega}$\,} % P_2Omega
\newcommand{\ps}{P$_{\omega}$\,} % P_omega
\newcommand{\ptwos}{P$_{2\omega}$\,} % P_2omega
\newcommand{\pb}{P$_{\omega-\Omega}$\,} % P_omega-Omega
\newcommand{\pbplus}{P$_{\omega+\Omega}$} % P_omega+Omega
\newcommand{\ptwob}{P$_{2(\omega-\Omega)}$} % P_2omega-Omega
\newcommand{\ptwobp}{P$_{2(\omega+\Omega)}$} % P_2omega+Omega
\newcommand{\pthreeb}{P$_{3(\omega-\Omega)}$} % P_3omega-Omega
\newcommand{\psone}{P$_{2\omega-\Omega}$} % P_2omega - Omega
\newcommand{\psfour}{P$_{\omega+2\Omega}$} % P_omega - 2Omega
\newcommand{\pnew}{P$_{2\Omega-\omega}$} % P_2Omega - omega
\title[A nearly synchronous IP]{SWIFT J0503.7-2819: A nearly synchronous intermediate polar below the period gap?}
\author[Rawat et al.]
{
Nikita Rawat$^{1,2}$\thanks{nikita@aries.res.in},
J. C. Pandey$^{1}$\thanks{jeewan@aries.res.in},
Arti Joshi$^{3}$, 
Simone Scaringi$^{4}$,
Umesh Yadava$^{2}$
\\
$^{1}$Aryabhatta Research Institute of Observational sciencES (ARIES), Nainital 263001, India\\
$^{2}$Deen Dayal Upadhyaya Gorakhpur University, Gorakhpur 273009, India\\
$^{3}$ Indian Institute of Astrophysics (IIA), Koramangala, Bangalore 560034, India\\
$^{4}$ Centre for Extragalactic Astronomy, Department of Physics, Durham University, South Road, Durham, DH1 3LE
}

\date{Accepted XXX. Received YYY; in original form ZZZ}

\pubyear{2022}

\begin{document}
\label{firstpage}
\pagerange{\pageref{firstpage}--\pageref{lastpage}}
\maketitle
\begin{abstract}
Based on the X-ray observations from \textit{XMM-Newton} and \textit{Swift},  and optical observations from Transiting Exoplanet Survey Satellite (\textit{TESS}) and AAVSO, we present temporal and spectral properties of probable intermediate polar SWIFT J0503.7-2819. The X-ray light curve shows two distinctive features, where possibly the second pole seems to be active during the middle of the \textit{XMM-Newton} observations. Present analysis confirms and also refines the previously reported orbital period of SWIFT J0503.7-2819 as 81.65$\pm$0.04 min. The X-ray and optical variations of this target have been found to occur at the period of $\sim$ 65 min, which we propose as the spin period of the white dwarf (WD). The energy-dependent modulation at this period, which are due to the photoelectric absorption in the accretion flow, also assures this conjecture. Two temperature thermal plasma model well explains the X-ray spectra with temperatures of  $\sim$ 150 eV and $\sim$ 18.5 keV, which is absorbed by a dense material with an average equivalent hydrogen column density of 3.8 $\times$ 10$^{22}$ cm$^{-2}$ that partially covers $\sim$ 27\% of the X-ray source. An attempt is made to understand the accretion flow in this system using the present data of SWIFT J0503.7-2819. If the proposed spin period is indeed the actual period, then SWIFT J0503.7-2819 could be the first nearly synchronous intermediate polar below the period gap.
\end{abstract}

\begin{keywords}
accretion, accretion discs, (stars:) cataclysmic variables, stars: individual: (SWIFT J0503.7-2819), stars: magnetic field 
\end{keywords}

\section{Introduction}
Intermediate Polars (IPs) are the low magnetic field strength (B $
\sim$ 10$^{6}$-10$^{7}$ G) subclass of Magnetic Cataclysmic Variables (MCVs). These are interacting binary systems containing a Roche-lobe filling late-type secondary star that transfers material to a magnetic primary WD \citep[see][for a full review of IPs]{1983ASSL..101..155W, 1994PASP..106..209P, 1995ASPC...85..185H}. The magnetic field of the WD plays a vital role in controlling the motion of the accretion flow within the effective magnetospheric radius. It also channels the accreting material towards the surface of the WD in the proximity of the magnetic poles. When the accreting matter reaches supersonic velocities, it forms a post-shock region above the WD surface. The post-shock region is hot ($\sim$ 10-80 keV) and the flow cools down via bremsstrahlung radiation emitting in the hard X-rays \citep{1995cvs..book.....W}  and cyclotron radiation \citep{1973PThPh..49.1184A}, which depends on the magnetic field strength of the WD. IPs are asynchronous systems and the majority of them follow the   asynchronism relation as \ps $\sim$ 0.1 \po, where \ps and \po are the spin period of WD and orbital period of the binary system, respectively. Further, the orbital period distribution of these systems shows the orbital periods longer than the `period gap' of 2-3 h \citep{2010MNRAS.401.2207S}. However, there is only one confirmed system, namely Paloma \citep{2007A&A...473..511S, 2016ApJ...830...56J, 2022arXiv220502863L, 2022arXiv220608635D}, which falls in the category of nearly synchronous IP, with \ps $\sim$ (0.7-0.9) \po. These systems are thought to be in the process of attaining synchronism and evolving into polars \citep{2004ApJ...614..349N}.

\par With the magnetic field strength of the WD <10 MG, an accretion disc can form, which is disrupted at the magnetospheric radius. Hence, the material is accreted either through an accretion disc or an accretion stream or a combination of both. Therefore, three accretion scenarios are believed to occur in IPs: disc-fed, stream-fed, and disc-overflow, depending upon the magnetic field strength of WD, mass accretion rate, and binary orbital separation. In the disc-fed accretion, the inner edge of the accretion disc is truncated at the magnetosphere radius, which results in the formation of `accretion curtains' near the magnetic poles of the WD \citep{1988MNRAS.231..549R}. A strong modulation at the spin frequency of the white dwarf \citep{1995A&A...298..165K, 1996MNRAS.280..937N} indicates the accretion to be occurring via a disc. In the disc-less or stream-fed accretion, the high magnetic field of the WD prevents the formation of a disc and infalling material is channelised along the magnetic field lines to the pole caps \citep{1986MNRAS.218..695H}. Modulation at the lower orbital sideband of the spin frequency, i.e. beat (\be) frequency \citep{1991MNRAS.251..693H, 1992MNRAS.255...83W, 1993MNRAS.265..316N} is a true indicator of stream-fed accretion. If there is an asymmetry between the magnetic poles, stream-fed accretion can also produce a modulation at the spin frequency, in addition to that at the beat frequency. \cite{1992MNRAS.255...83W} showed the importance of \sone frequency in distinguishing between these two modes of accretion in hard X-ray regimes. The \sone frequency is only present in disc-less systems along with sometimes dominant \orb component, \be, and \s. The disc-overflow accretion \citep{1989ApJ...340.1064L, 1996ApJ...470.1024A}, where both disc-fed and stream-fed accretions can occur as an accretion disc is present, but a part of the accretion stream skips the disc and directly interacts with the WD magnetosphere \citep{1989MNRAS.238.1107H, 1991ApJ...378..674K}. For a disc-overflow accretion, modulations at both \s and \be frequencies are expected to occur \citep[see][]{1991MNRAS.251..693H, 1993MNRAS.265L..35H}. 

\par SWIFT J0503.7-2819 (hereafter J0503) was found to be a variable X-ray source in the 
%Neil Gehrels Swift Observatory (henceforth 
\textit{Swift}-XRT images of the field and UV bright star in the \textit{Swift} UVOT by \citet{2015AJ....150..170H}. It is located at a distance of 837$_{-43}^{+60}$ pc \citep{2021AJ....161..147B}. The optical spectrum of J0503 showed typical features of a CV: strong Balmer, He \Romannum{1}, and He \Romannum{2} lines on a blue continuum, with He \Romannum{2} $\lambda$4686 roughly equal in strength to H$\rm \beta$ \citep{2015AJ....150..170H}. The value of equivalent width (EW) ratio He \Romannum{2}/H$\rm \beta$ between 0.5 and 1 suggest that J0503 could be an IP \citep{2019Ap&SS.364..153M}. 
From radial-velocity periodogram analysis and time-series photometric analysis, \citet{2015AJ....150..170H} also derived the orbital period of the system to be 81.60(7) min. Although \cite{2022ApJ...934..123H} discusses that their derived spectroscopic period might not necessarily be the true orbital period. \citet{2015AJ....150..170H} also found a signal at 975.2 s, which was provisionally suggested to be the spin period of the WD. J0503 is classified as a probable IP in the IP catalogue of Koji Mukai\footnote{\url{https://asd.gsfc.nasa.gov/Koji.Mukai/iphome/catalog/alpha.html}}.

\par During the final preparation of this manuscript, \citet{2022ApJ...934..123H} published the timing properties of the source using \textit{XMM-Newton} and \textit{TESS} data sets, which we have also used in the present study along with AAVSO and \textit{Swift} observations. The significant difference presented here and in \citet{2022ApJ...934..123H} is that the author did not study the X-ray spectral properties, which is generally considered a powerful tool to understand the physical properties of the mass accretion flow in these accreting systems. Further, the timing analysis presented here is slightly different from that done by \citet{2022ApJ...934..123H}. The paper is structured as follows. Section \ref{sec2} describes the observations and data used for this study. Section \ref{sec3} contains our analysis and results. Finally, the discussion and conclusions are presented in Sections \ref{sec4} and \ref{sec5}, respectively.

\section{Observations and Data Reduction} \label{sec2}

\subsection{\textit{XMM-Newton} Observations}\label{sec2.1}
J0503 was observed by the \textit{XMM-Newton} satellite \citep{2001A&A...365L...1J} using the European Photon Imaging Camera \citep[EPIC;][]{2001A&A...365L..18S, 2001A&A...365L..27T} on 2018 March 7 at 10:45:51 (UT) with an offset of 0.012 arcmin (observation ID: 0801780301). The exposure times for the p-n junction (PN) and metal oxide semiconductor (MOS) detectors were 26.6 ks and 25 ks, respectively. We have used the standard \textit{XMM-Newton} Science Analysis System (SAS) software package (version 20.0.0) with the latest calibration files at the time of analysis\footnote{\url{https://www.cosmos.esa.int/web/xmm-newton/current-calibration-files}} for the data reduction. We have followed the SAS analysis thread\footnote{\url{https://www.cosmos.esa.int/web/xmm-newton/sas-threads}} for data reduction and used SAS tools \textit{epproc} and \textit{emproc} to produce calibrated event files. We have also corrected event arrival times to the solar system barycenter with the \textit{barycen} task.  We have inspected the data for the high background proton flares and found that these were free from this effect. We have also checked the existence of pile-up using the \textit{epatplot} task but did not find any significant presence of it. To avoid the background contribution at higher energies, we have carried out further analyses in the energy range of 0.3-10.0 keV. We have chosen a circular source region with a 30 arcsec radius centring the source and a circular background region with a similar size to that of the source from the same CCD to extract the final light curve, spectrum, and detector response files. The spectra have been rebinned with the \textit{grppha} tool to minimum 20 counts per bin. Further, temporal and spectral analyses were done using HEASOFT version 6.29.
\par J0503 was also observed using the optical monitor \citep[OM;][]{2001A&A...365L..36M} in the V filter for a total exposure time of 25.4 ks. OM fast mode data were reprocessed with the task \textit{omfchain} and the individual exposures were merged together to get the summed light curve file.

\subsection{\textit{Swift} Observations} \label{sec2.2}
J0503 was first observed by the \textit{Swift} on two occasions, 2010 June 2 at 11:28:00 (UT) and 2010 June 9 at 04:28:59 (UT) using X-ray Telescope \citep[XRT;][]{2005SSRv..120..165B} with offsets of 4.1 arcmin (observation ID: 00041156001) and 4.7 arcmin (observation ID: 00041156002), respectively. The exposure times for two IDs were 6271 s and 1782 s, respectively. The XRT observes in the 0.3-10.0 keV energy range. The task \textit{xrtpipeline} (version 0.13.6) along with the latest calibration files were used to produce the cleaned and calibrated event files. The barycentric correction was applied to both event files using the task \textit{barycorr}. The source light curves and spectra were extracted by selecting a circular region of 22 arcsec radius. The background was chosen from a nearby source-free region with a 50 arcsec radius. An exposure map was built using the task \textit{xrtexpomap} to correct for the loss of flux caused by some of the CCD pixels not being used to collect data. The output of this task was then used to make an ancillary response file to correct the loss of the counts due to hot columns and bad pixels with the task \textit{xrtmkarf}. We have used the response matrix file, swxpc0to12s6\_20090101v014.rmf, provided by the \textit{Swift} team. Both spectra from the XRT were rebinned using the \textit{grppha} for a minimum of one count per bin and spectral fits were performed using the C-statistics.

\subsection{\textit{TESS}, ASAS-SN, and AAVSO Observations}\label{sec2.3}
The \textit{TESS} observations of J0503 were carried out during sector 32 from 2020 November 19 to 2020 December 16 at a cadence of 2 min. The total observing time was $\sim$ 26 d with a small gap of $\sim$ 1.66 d in the middle due to the downlink of data at perigee. The \textit{TESS} bandpass extends from 600 to 1000 nm with an effective wavelength of 800 nm \citep[see][for details]{2015JATIS...1a4003R}. The data were available at Mikulski Archive for Space Telescopes (MAST) data archive\footnote{\url{https://mast.stsci.edu/portal/Mashup/Clients/Mast/Portal.html}} with a unique identification number `686160823'. The \textit{TESS} pipeline provides two flux values: simple aperture photometry (SAP) and pre-search data conditioned SAP (PDCSAP). The PDCSAP light curve attempts to remove instrumental systematic variations by fitting and removing those signals that are common to all stars on the same CCD\footnote{see section 2.1 of \textit{TESS} archive manual available at \url{https://outerspace.stsci.edu/display/TESS/2.0+-+Data+Product+Overview}}. In this way, aperiodic variability  might be removed from the data. PDCSAP also corrects for the amount of flux captured by the photometric aperture and crowding from known nearby stars, while SAP does not\footnote{see section 2.0 of \textit{TESS} archive manual available at \url{https://outerspace.stsci.edu/display/TESS/2.0+-+Data+Product+Overview}}. In order to assure that any aperiodic variability has not been  removed from PDCSAP data, we have performed the periodogram analysis of both SAP and PDCSAP data. Both SAP and PDCSAP data resulted in similar power spectra indicating that PDCSAP data of J0503 can be safely used for further analysis. Therefore, we have considered PDCSAP flux values with the `quality flag' value of 0.
\par We have also utilized the publicly available V-band data of J0503 from All-Sky Automated Survey for Supernovae \citep[ASAS-SN\footnote{\url{https://asas-sn.osu.edu/variables}};][]{2014ApJ...788...48S, 2017PASP..129j4502K} and American Association of Variable Star Observers \citep[AAVSO\footnote{\url{https:// www.aavso.org/}};][]{2021.K} CV-band (unfiltered data with a V-band zero-point) data to represent the long-term variability of the source. The unavailability of simultaneous data points between \textit{TESS} and ground-based data did not provide us with any means to compare these data sets.

\section{Data Analysis and Results }\label{sec3}
\subsection{Timing Analysis}\label{sec3.1}
\subsubsection{X-ray Light Curves and Power Spectra} \label{sec3.11}
\begin{figure}
\centering
\includegraphics[width=8.5cm, height=6cm]{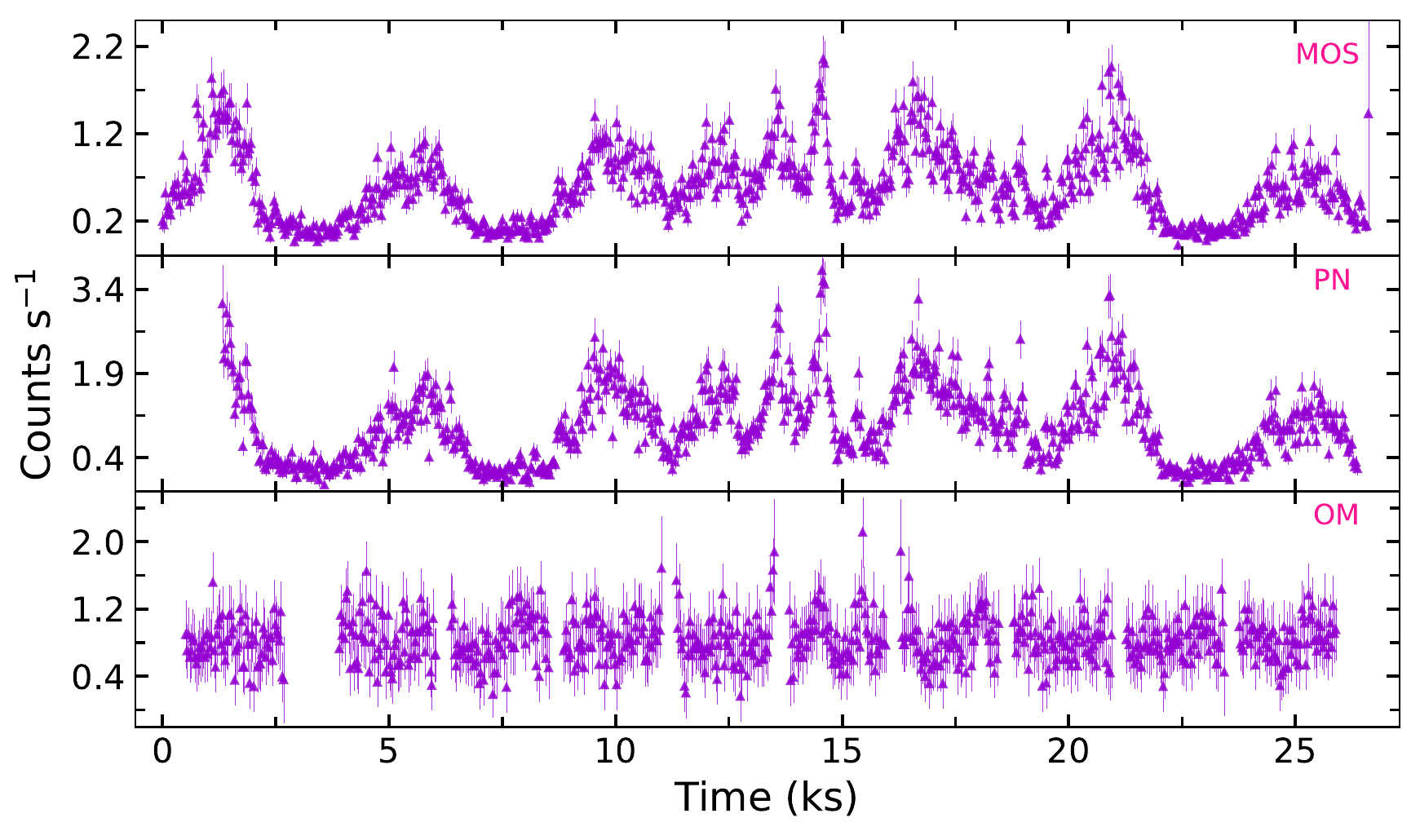}
\caption{EPIC light curves in the 0.3–10.0 keV energy range and OM light curve in V filter binned in 30 s intervals. }
\label{fig:lc_epic}
\end{figure}

The background-subtracted X-ray light curves of J0503 obtained from the \textit{XMM-Newton} observations in the 0.3-10.0 keV energy band are shown in the top two panels of Figure \ref{fig:lc_epic}. The temporal binning of 30 s was used to extract the light curve for MOS and PN detectors. Both light curves show continuous and periodic intensity variations. If we closely inspect both light curves, we see two distinctive features: broad minima, which cover two different time spans (i.e. up to 10 ks from the start of observations and 22 ks to the end of observations) and no broad minima between 10-22 ks of X-ray observations. \citet{2022ApJ...934..123H} suggested that these broad minima  are due to the  self-eclipse of a single active accretion region, whereas the middle part of the light curves shows that a second accretion region (possibly, the second pole) becomes active during this duration. Therefore, we  refer to broad minima timings of the X-ray light curve as  accretion from pole-1 whereas the middle timings of the X-ray light curve as accretion from pole-2.

% Figure-1

\par To find the periodicities in the data, we have performed Lomb-Scargle (LS) periodogram analysis \citep{1976Ap&SS..39..447L, 1982ApJ...263..835S}. The top panels of Figures \ref{fig:mos_ps} and \ref{fig:pn_ps} show LS power spectra in the 0.3-10.0 keV energy band. The energy-resolved power spectra in 0.3-1.0 keV, 1.0-2.0 keV, 2.0-5.0 keV, and 5.0-10.0 keV are also shown in the lower panels of Figures \ref{fig:mos_ps} and \ref{fig:pn_ps}. The periods corresponding to the signiﬁcant peaks in the power spectra of the 0.3-10.0 keV energy band are given in Table \ref{tab:periods}. The signiﬁcance of these detected peaks was determined by calculating the false alarm probability \citep[FAP;][]{1986ApJ...302..757H}. The horizontal dashed line in each power spectrum represents the 90\% confidence level. If the light curve has intrinsic red noise variability, then FAP needs to be considered as a conservative limit. The dominant peak corresponds to a period of $\sim$ 65 min, which we assign as the spin period of the WD. The period of $\sim$ 84 min was identified as the orbital period of the binary system and consistent with the earlier finding \citep{2015AJ....150..170H}. We have inferred the beat period  (\pb)  of $\sim$ 332 min by using the precisely determined  values of \ps and \po from the power spectral analysis of the longer spanned  \textit{TESS} light curve (see section \ref{sec3.12} for details). The beat period obtained from MOS and PN is well within a 1$\sigma$ level of this inferred value. The other derived periods can be identified as  \ptwoo, \ptwos, P$_{3\omega}$, P$_{4\omega}$, \ptwob, \pthreeb, \pbplus, \ptwobp,  \psfour, and \psone. We have detected \pb, \ptwob, and \pthreeb in the power spectra, unlike \citet{2022ApJ...934..123H}. The reason for this could be either combining both MOS and PN data sets or a large temporal binning in his work.  Further, the inferred value of \psone ~comes out to be $\sim$ 55 min, which closely matches the period we have obtained in the X-ray power spectra (see Table \ref{tab:periods} and Figure \ref{fig:epic_ps}). 
\par
\begin{figure}
\centering
\includegraphics[width=8cm, height=3.5cm]{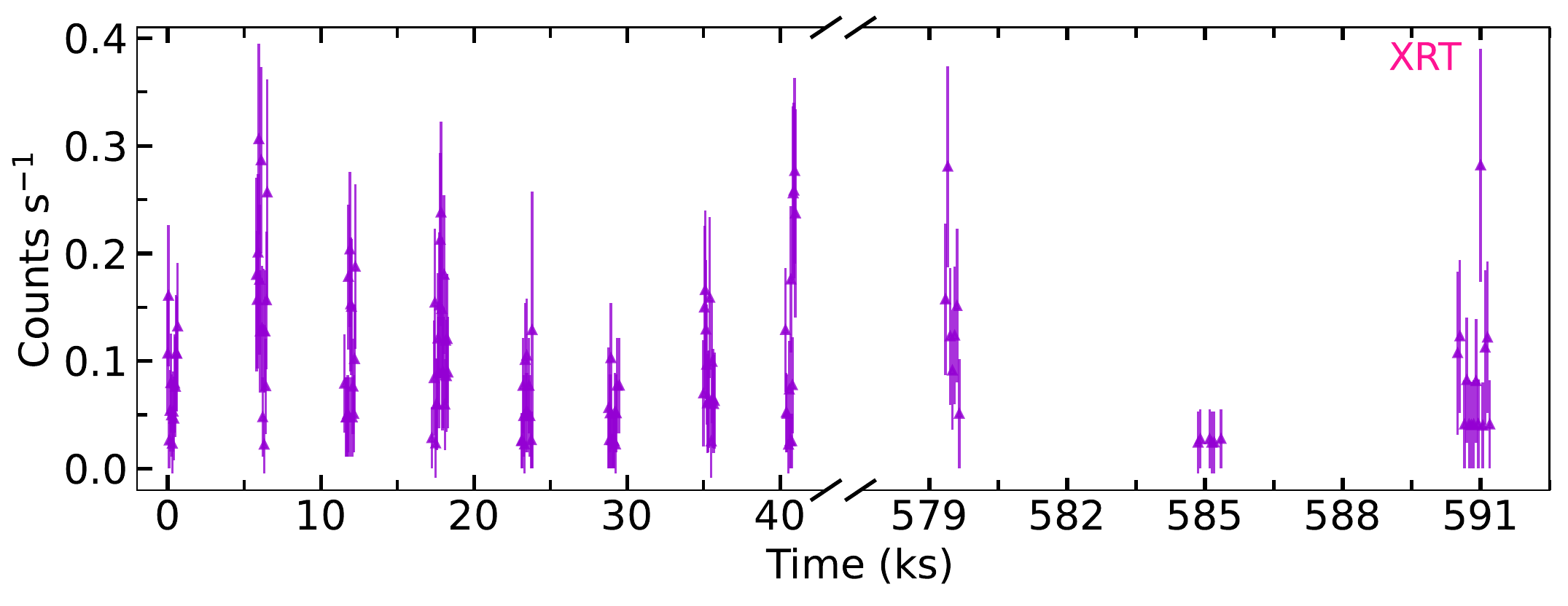}
\caption{\textit{Swift}-XRT light curves in the 0.3–10.0 keV energy range binned in 50 s intervals. }
\label{fig:lc_sw}
\end{figure}

\begin{figure*}
\centering
\subfigure[]{\includegraphics[width=7cm, height=8cm]{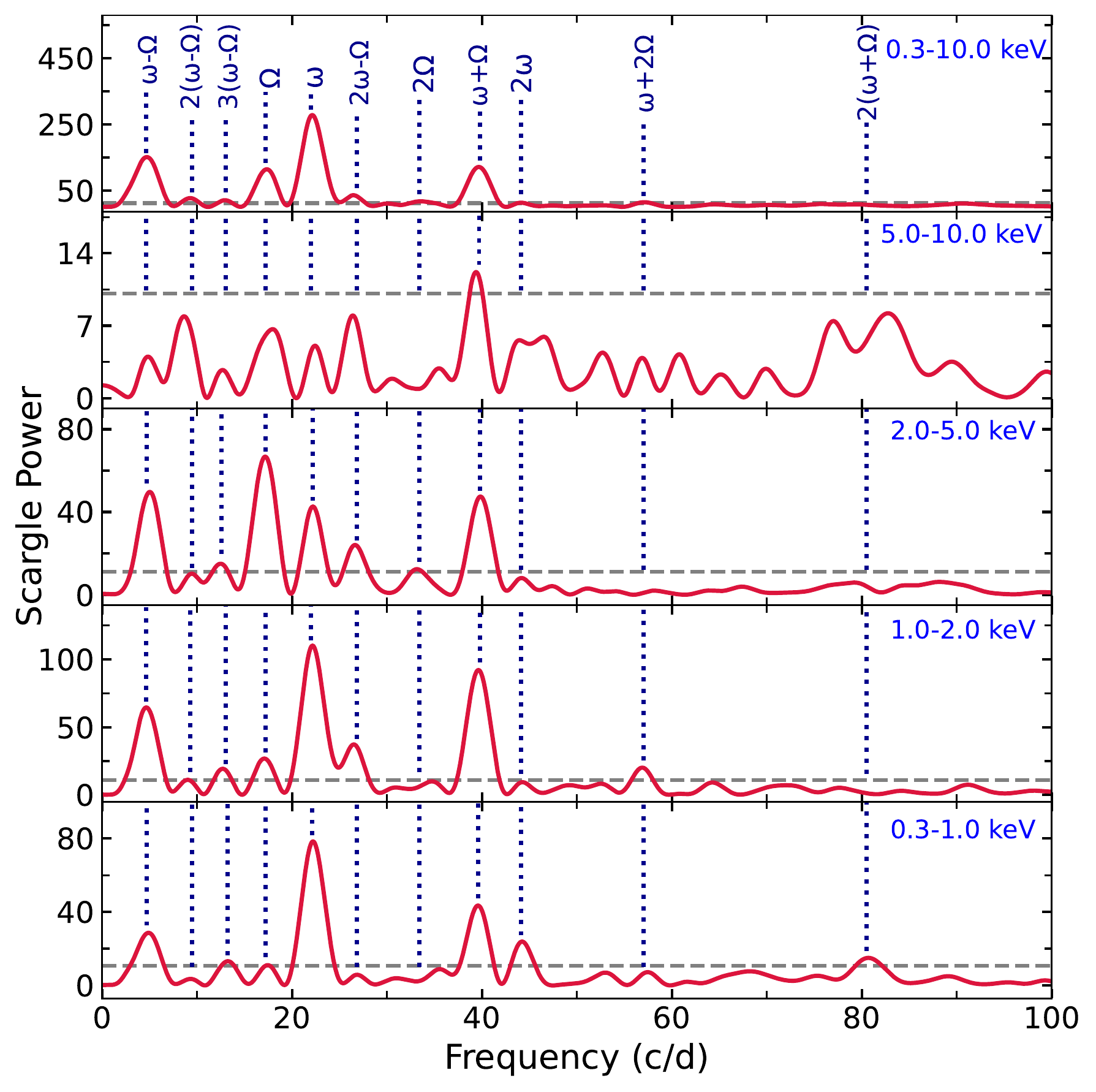} \label{fig:mos_ps}}
\subfigure[]{\includegraphics[width=7cm, height=8cm]{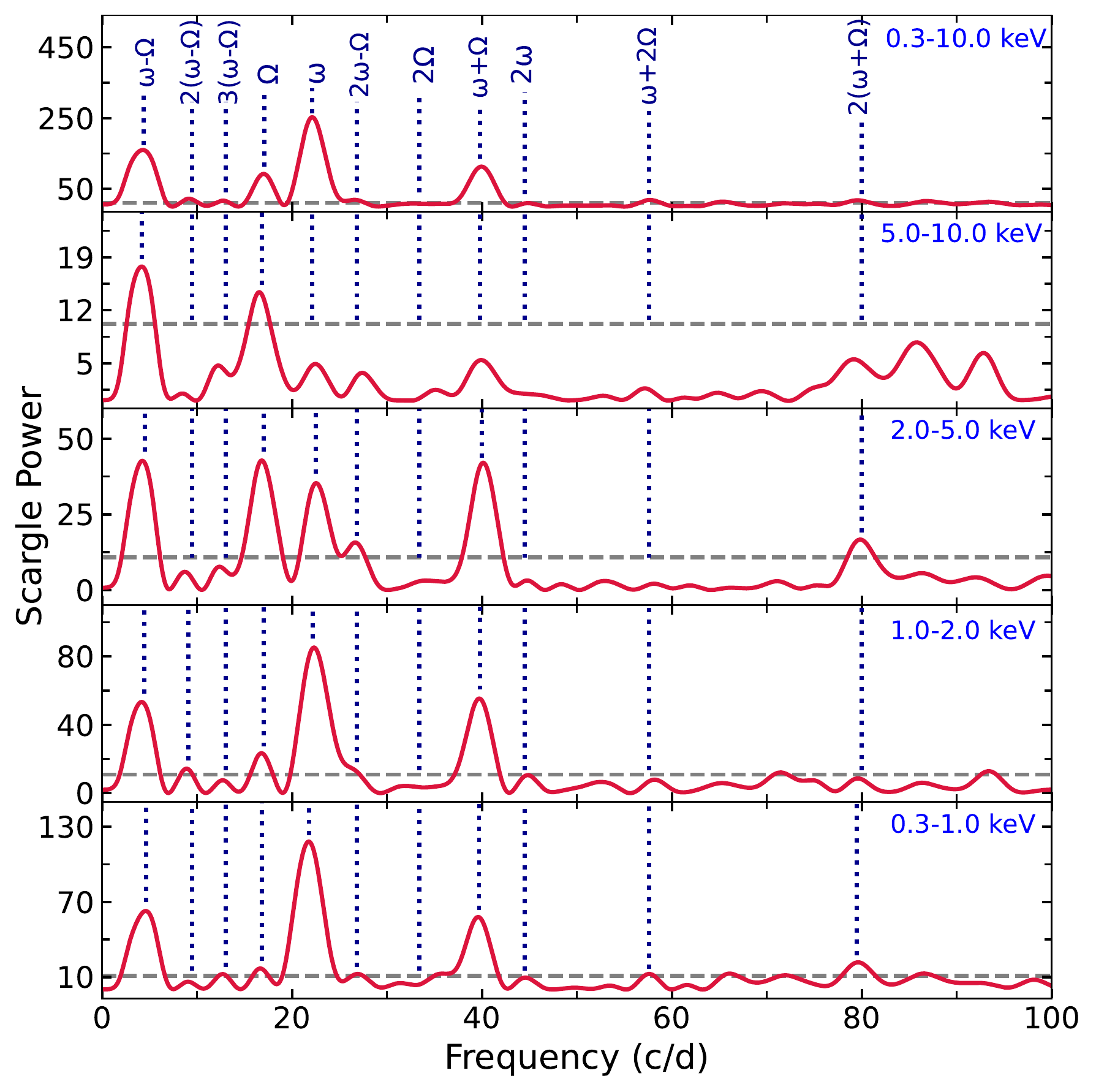} 
\label{fig:pn_ps}}
\caption{Power spectra of J0503 obtained from (a) EPIC-MOS and (b) EPIC-PN data in the different energy bands. From top to bottom, the panels show the X-ray power spectra in 0.3-10.0 keV, 5.0-10.0 keV, 2.0-5.0 keV, 1.0-2.0 keV, and 0.3-1.0 keV energy bands, respectively. The grey horizontal dashed line in each panel represents the 90\% confidence level. Significant frequencies are marked for clear visual inspection.}
\label{fig:epic_ps}
\end{figure*}

 Figure \ref{fig:lc_sw} shows background-subtracted X-ray light curves of J0503 in the 0.3-10.0 keV energy band as obtained from the two observations of \textit{Swift-XRT}. The sparse data points resulted in a very noisy power spectrum and the real frequencies were found to be hidden under the noise. Therefore, we have not shown the power spectrum obtained from the XRT data.

% Figure-2

% Figure-3

% Table-1
\begin{table*}
\begin{center}
\caption{Periods corresponding to the dominant frequency peaks in the power spectra of J0503 obtained from X-ray (0.3-10.0 keV) and optical (\textit{TESS}, OM, and AAVSO) data.}
\label{tab:periods}
\end{center}
\renewcommand{\arraystretch}{1.4}
\begin{tabular}{lccccc}
\hline
\multirow{2}{*}{Identification} & \multicolumn{5}{c}{Period (minutes)}\\
\cline{2-6}
& MOS & PN & OM & \textit{TESS} & AAVSO\\
 \hline
\po & 84.5 $\pm$ 4.0 & 83.4 $\pm$ 4.2 &  --- & 81.65 $\pm$ 0.04 & 81.63  $\pm$ 0.02\\
\ptwoo & 43.3 $\pm$ 1.1 & --- & 41.29 $\pm$ 1.01 & 40.81 $\pm$ 0.01 &   40.81 $\pm$ 0.01\\
\ps & 65.7 $\pm$ 2.4 & 64.2 $\pm$ 2.5 & --- & 65.53 $\pm$ 0.03  & 65.53 $\pm$ 0.01\\
\ptwos & 32.8 $\pm$ 0.6 & --- & --- & --- & ---\\
P$_{3\omega}$ & --- & 21.9 $\pm$ 0.3 & --- & --- & ---\\
P$_{4\omega}$ & --- & 16.5 $\pm$ 0.2 & --- & --- & ---\\
\pb & 295.7 $\pm$ 50.7 & 333.7 $\pm$ 69.5 & --- & --- & ---\\
\ptwob & 161.3 $\pm$ 14.8 & 151.7 $\pm$ 13.9 & --- & --- & ---\\
\pthreeb & 110.9 $\pm$ 6.9 & 111.2 $\pm$ 7.4 & --- & --- & ---\\
\pnew & --- & --- & --- & 108.22 $\pm$ 0.08  & 108.14 $\pm$ 0.03\\
\pbplus & 36.2 $\pm$ 0.7 & 36.3 $\pm$ 0.8 & --- & --- & ---\\
\ptwobp & --- & 18.1 $\pm$ 0.2 & --- & --- & ---\\
\psfour & 25.3 $\pm$ 0.4 & 24.9 $\pm$ 0.4 & --- & --- & ---\\
\psone & 53.8 $\pm$ 1.6 & --- & --- & --- & ---\\
\hline
\end{tabular}
\end{table*}

% Figure-4
\begin{figure*}
%\centering
\includegraphics[width=15cm, height=4.5cm]{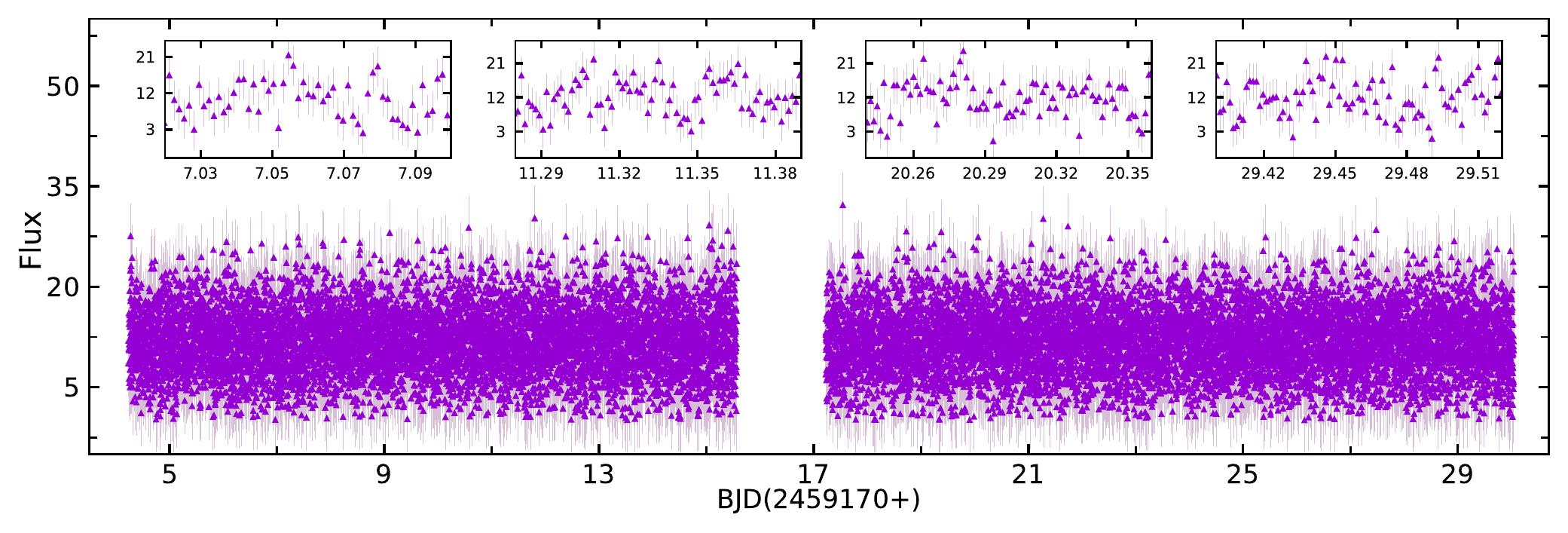}
\caption{\textit{TESS} light curve of J0503. Inset of the figure shows a close-up of some variability cycles. }
\label{fig:lc_tess}
\end{figure*}

% Figure-5
\begin{figure*}
%\centering
\includegraphics[width=15cm, height=4.5cm]{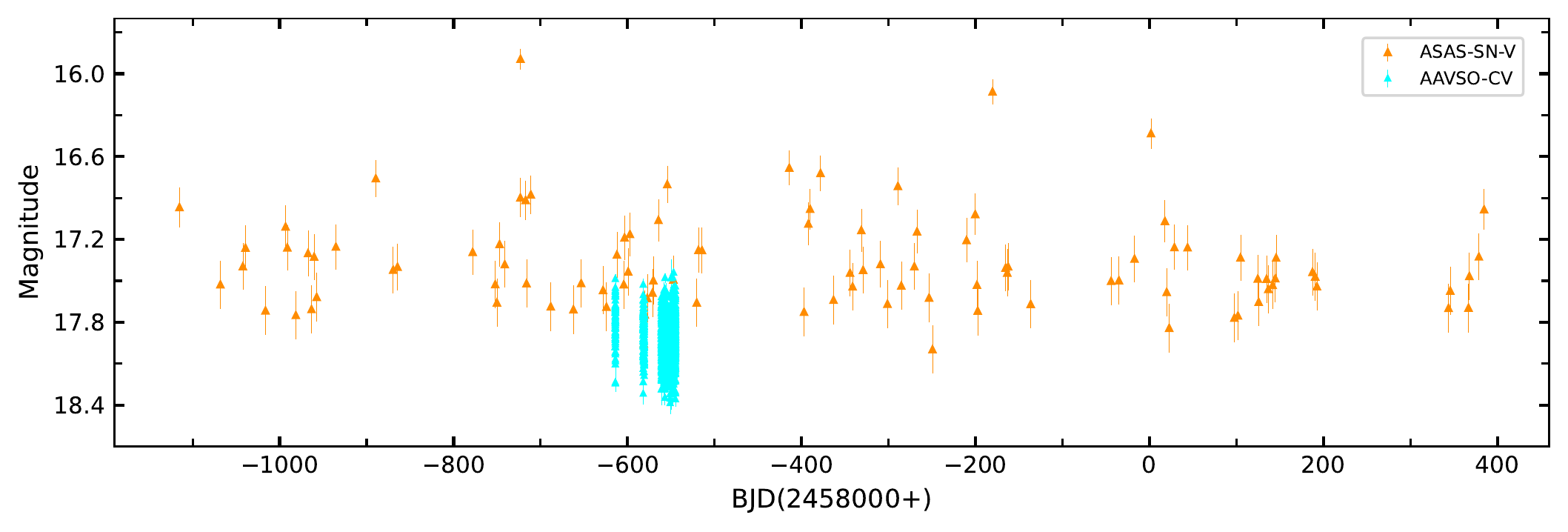}
\caption{Long-term variable light curve of J0503 using ASAS-SN and AAVSO observations. }
\label{fig:lc_all}
\end{figure*}

% Figure-6
\begin{figure*}
\centering
\includegraphics[width=15cm, height=8cm]{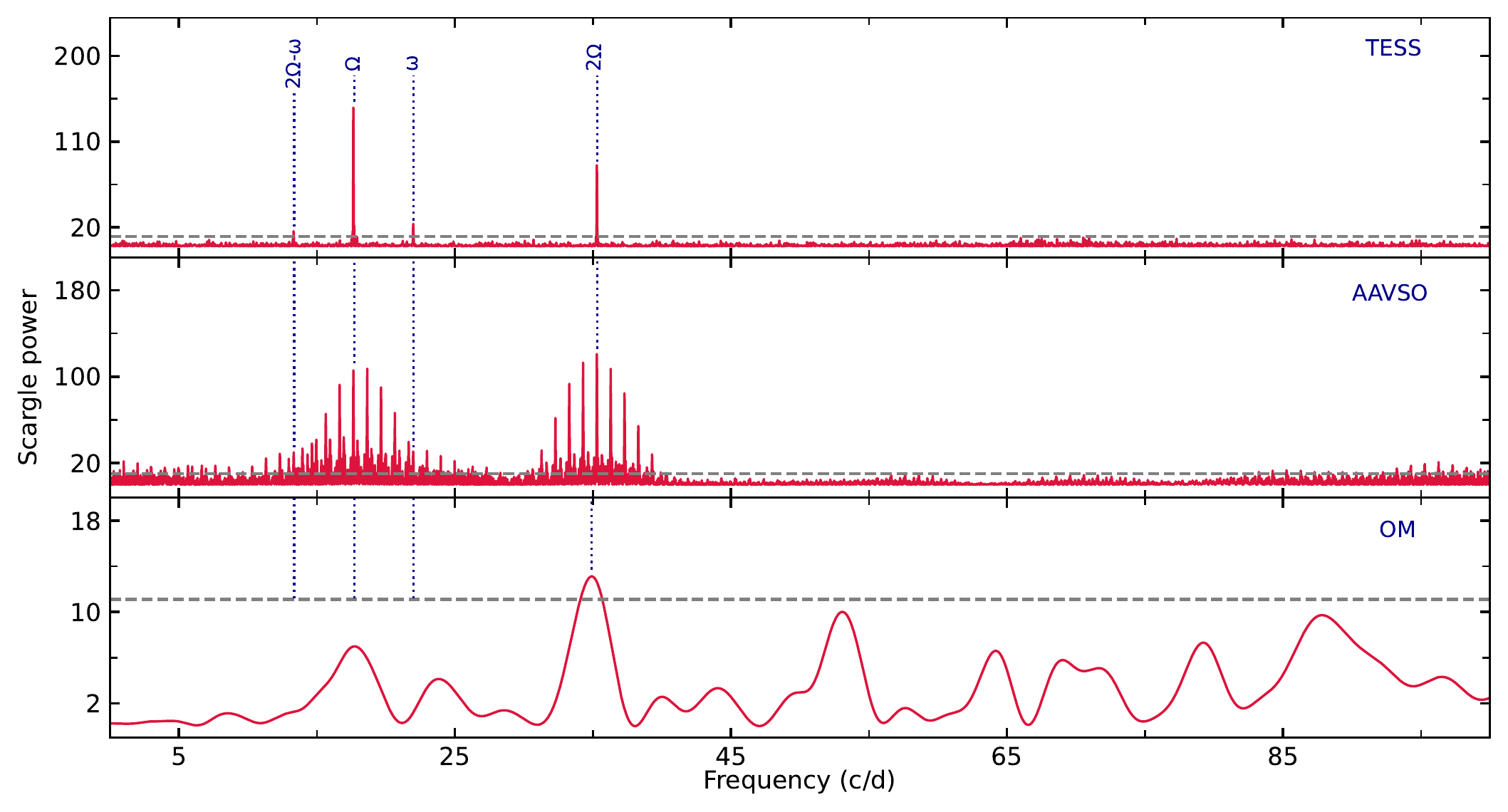}
\caption{Optical power spectra of J0503 as obtained from \textit{TESS}, AAVSO, and OM. The grey horizontal dashed line represents the 90\% confidence level. Major frequencies are marked for clear visual inspection.}
\label{fig:ps_all}
\end{figure*}

% Figure-7
\begin{figure*}
\centering
\subfigure[]{\includegraphics[width=7cm, height=8cm]{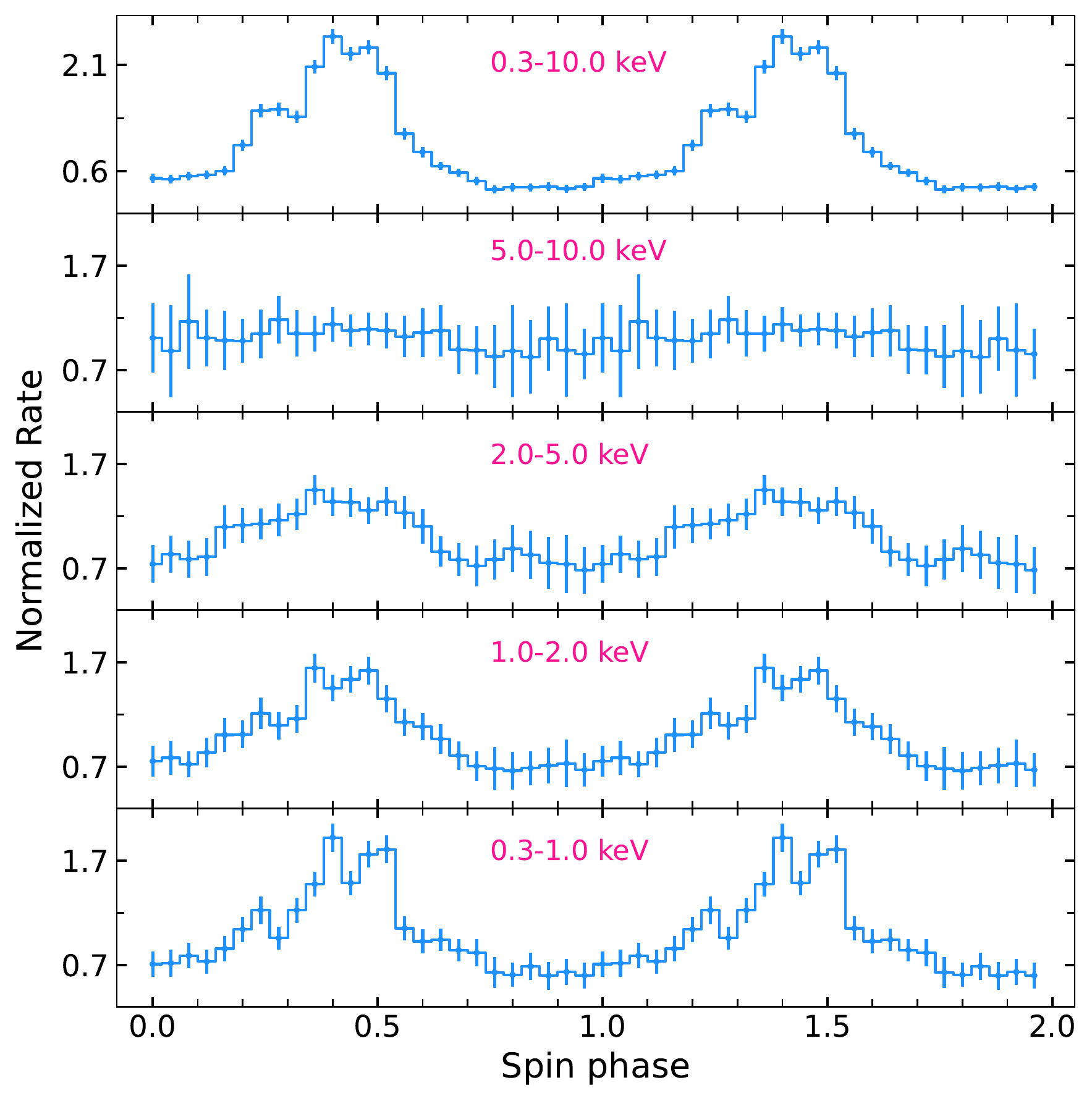} \label{fig:mos_spin}}
\subfigure[]{\includegraphics[width=7cm, height=8cm]{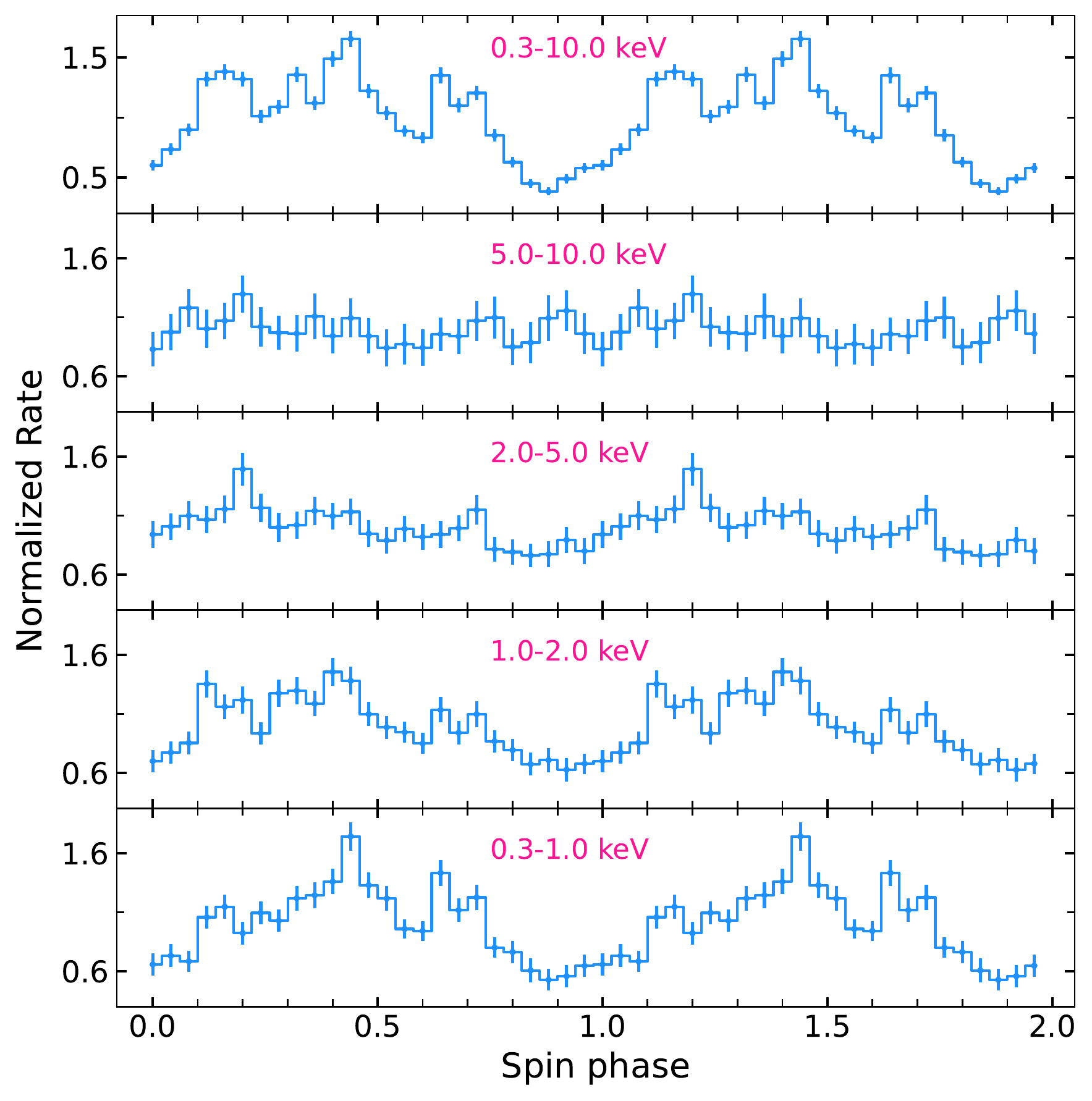} 
\label{fig:pn_spin}}
\caption{Energy-dependent folded light curves of (a) pole-1 and (b) pole-2 data at the spin period of 65.53 min. }
\label{fig:spin-fold}
\end{figure*}

% Figure-8
\begin{figure*}
\centering
\subfigure[]{\includegraphics[width=7cm, height=8cm]{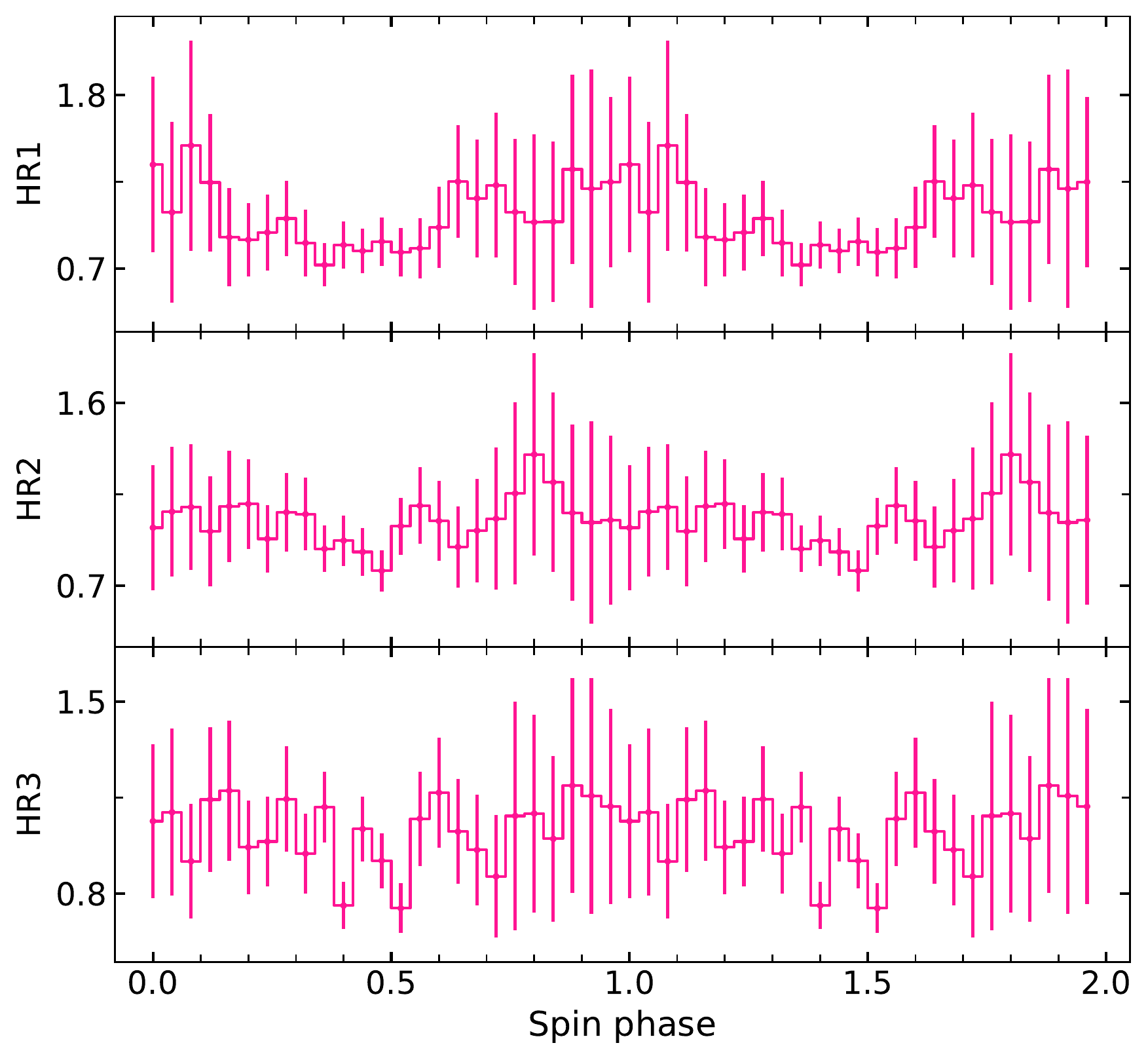} \label{fig:pn-hr-pole1}}
\subfigure[]{\includegraphics[width=7cm, height=8cm]{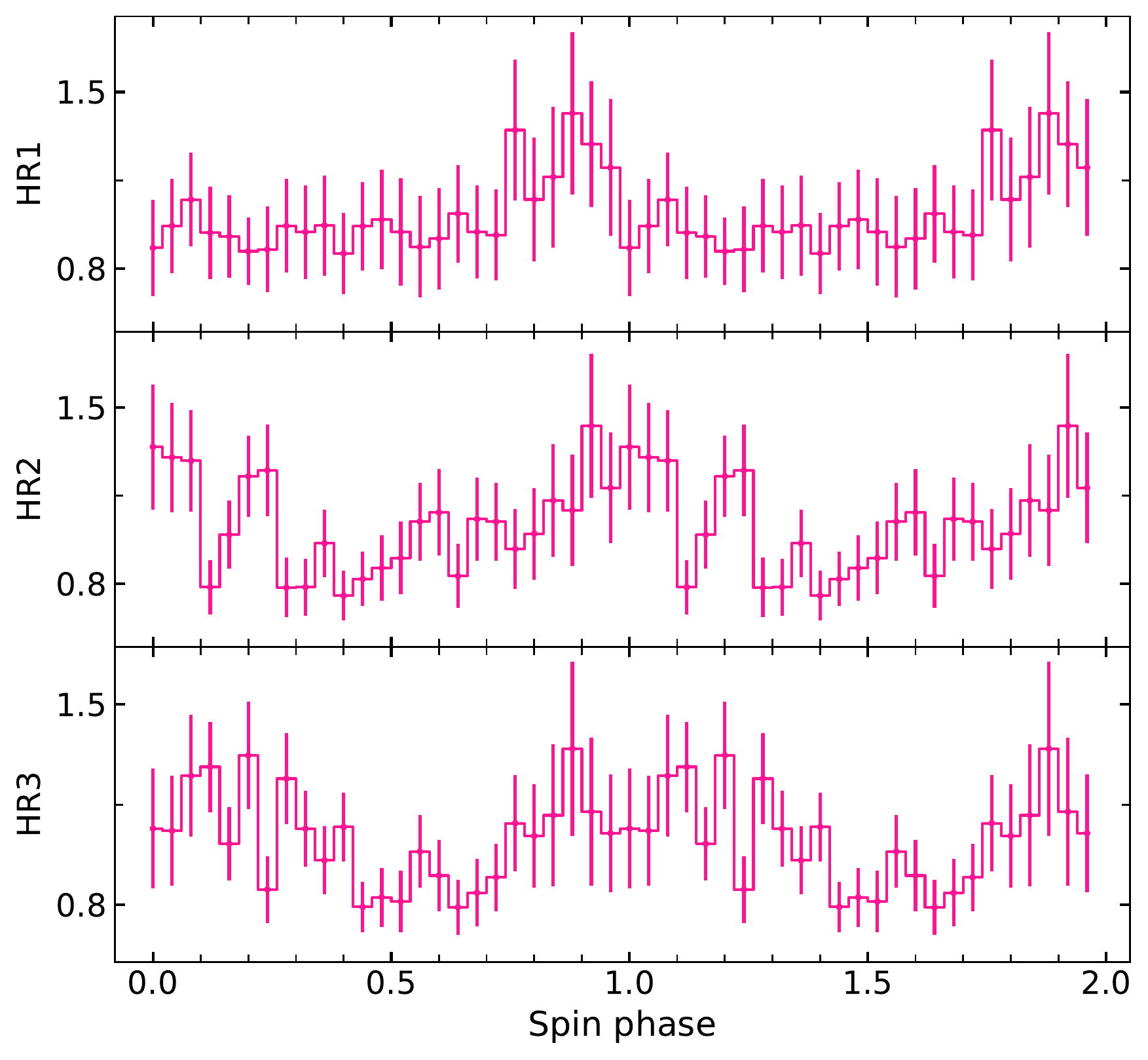} 
\label{fig:pn-hr-pole2}}
\caption{ Hardness ratio curves HR1, HR2, and HR3 of (a) pole-1 and (b) pole-2 data where HR1 is the ratio of the count rate in 5.0-10.0 keV to the count rate in the 2.0-5.0 keV energy bands, i.e., HR1=(5-10)/(2-5), HR2 is the ratio of the count rate in 2.0-5.0 keV to the count rate in 1.0-2.0 keV energy bands, i.e., HR2=(2-5)/(1-2), and HR3 is the ratio of the count rate in 1.0-2.0 keV to the count rate in 0.3-1.0 keV energy bands, i.e., HR3=(1-2)/(0.3-1).}
\label{fig:hr-spin}
\end{figure*}

% Figure-9
\begin{figure*}
\centering
\subfigure[]{\includegraphics[width=7cm, height=8cm]{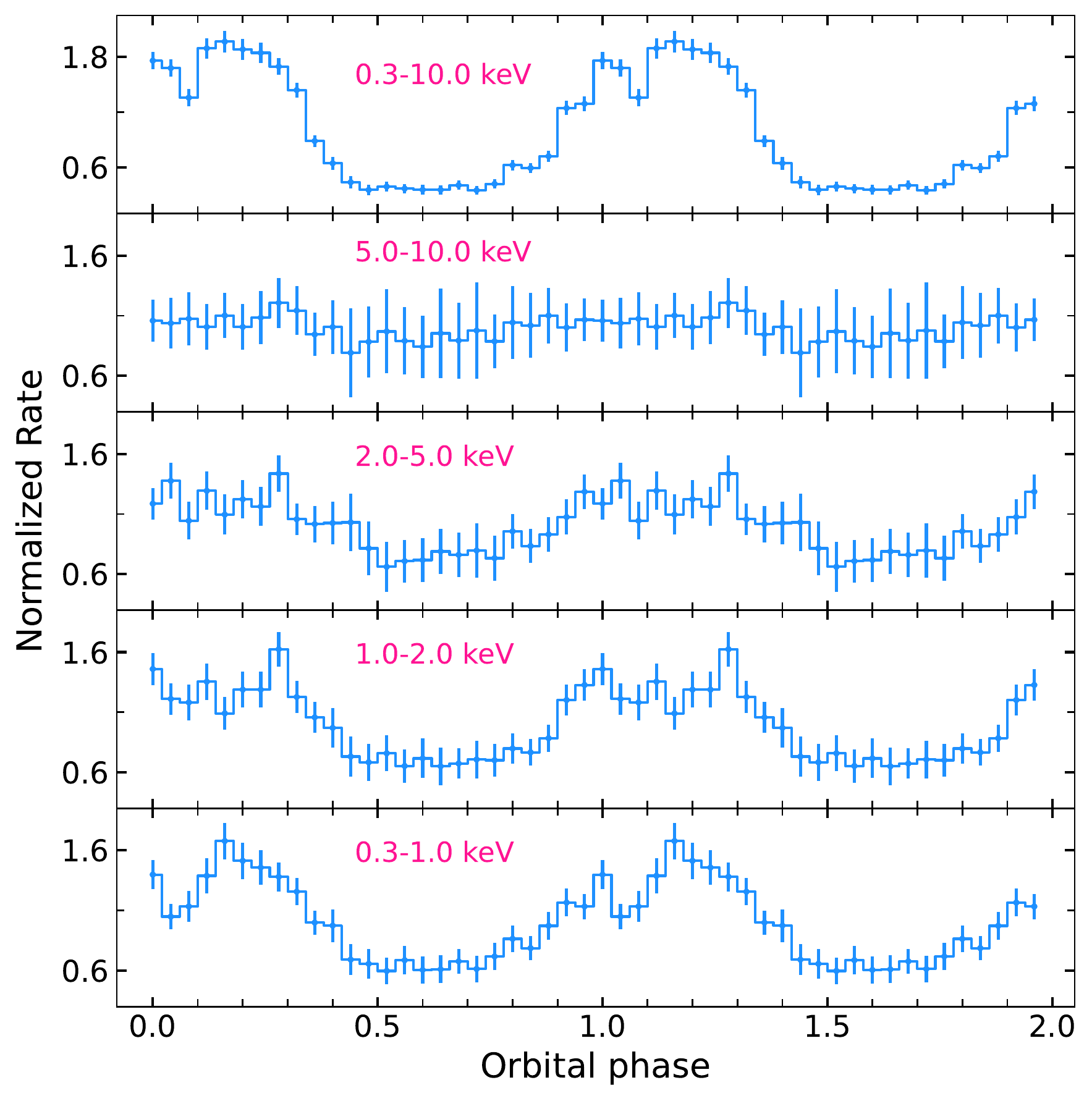} \label{fig:mos_orb}}
\subfigure[]{\includegraphics[width=7cm, height=8cm]{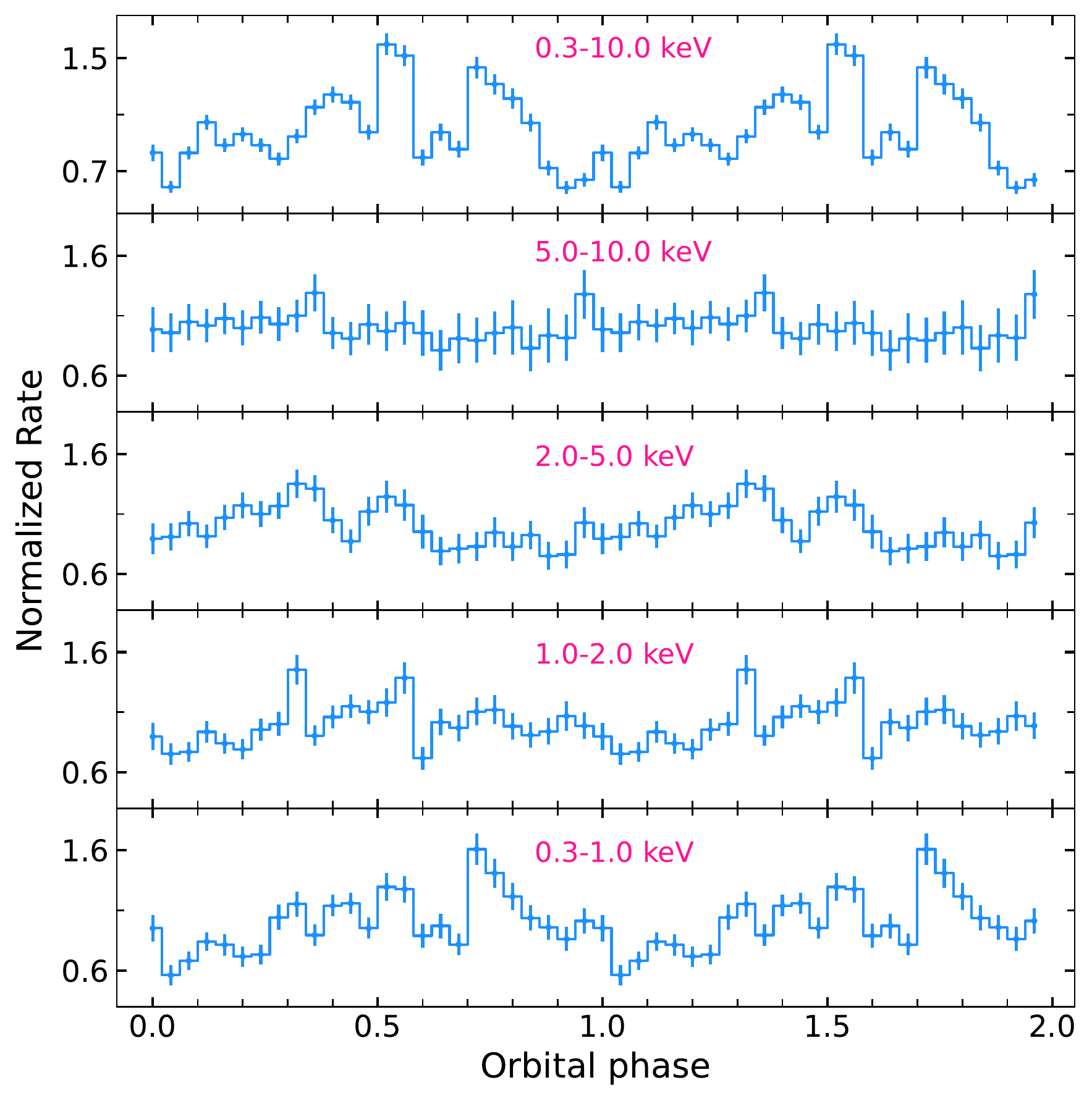} 
\label{fig:pn_orb}}
\caption{Energy-dependent folded light curves of (a) pole-1 and (b) pole-2 data at the orbital period of 81.65 min.}
\label{fig:orb-fold}
\end{figure*}

% Figure-10
\begin{figure*}
\centering
\subfigure[]{\includegraphics[width=7cm, height=8cm]{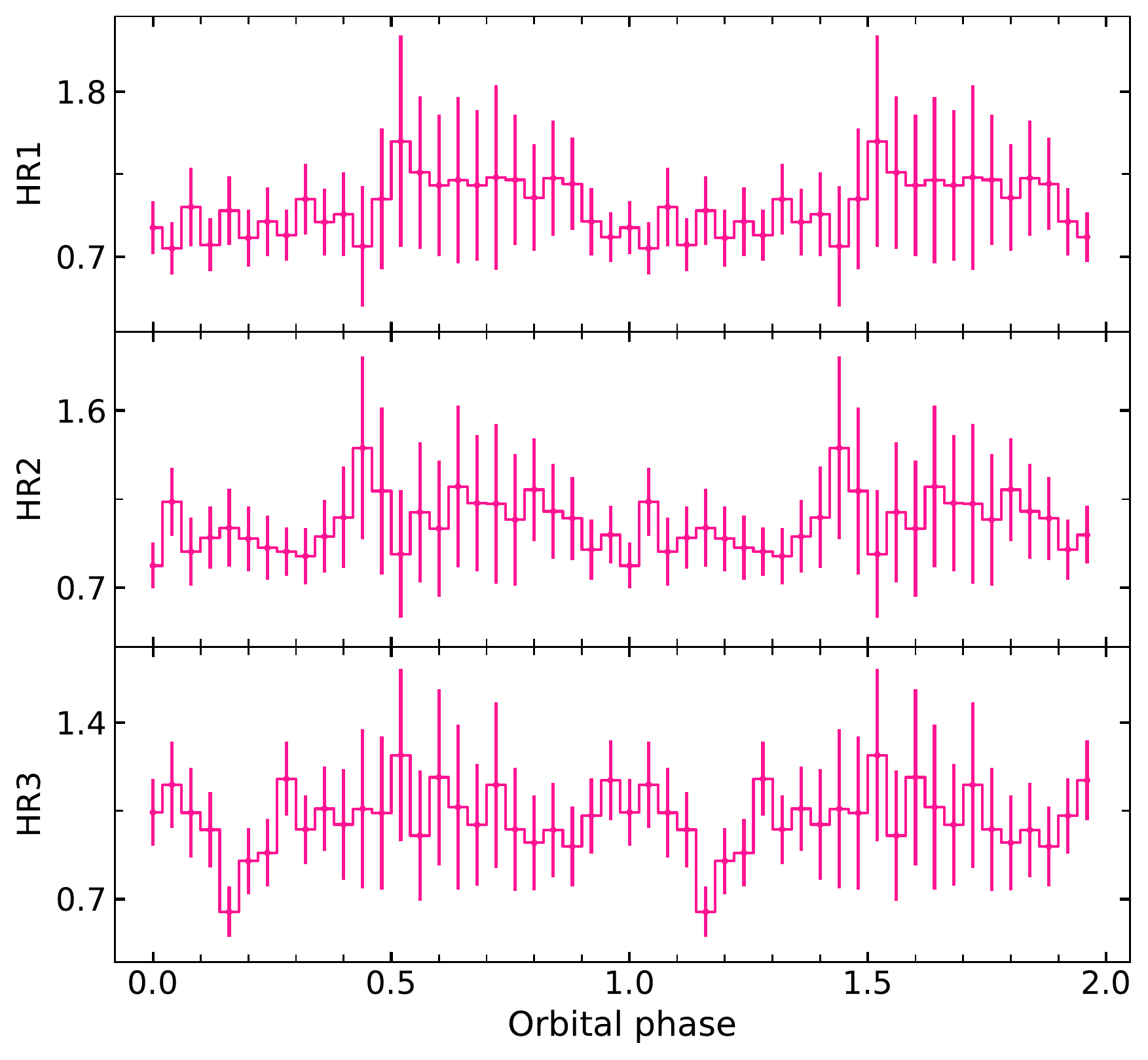} \label{fig:hr-pole1-orb}}
\subfigure[]{\includegraphics[width=7cm, height=8cm]{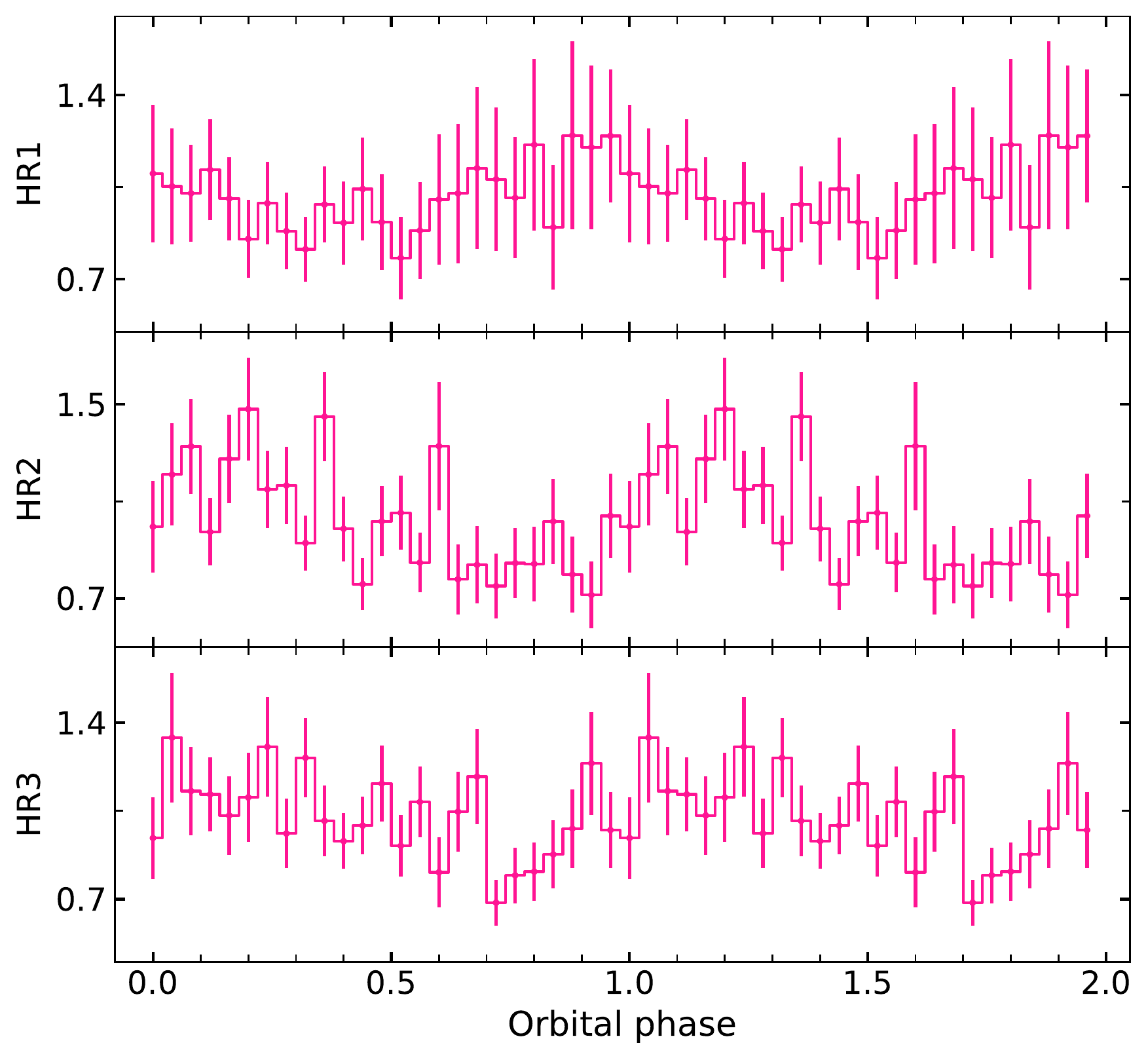}
\label{fig:hr-pole2-orb}}
\caption{ Hardness ratio curves HR1, HR2, and HR3 of (a) pole-1 and (b) pole-2 data where HR1 is the ratio of the count rate in 5.0-10.0 keV to the count rate in the 2.0-5.0 keV energy bands, i.e., HR1=(5-10)/(2-5), HR2 is the ratio of the count rate in 2.0-5.0 keV to the count rate in 1.0-2.0 keV energy bands, i.e., HR2=(2-5)/(1-2), and HR3 is the ratio of the count rate in 1.0-2.0 keV to the count rate in 0.3-1.0 keV energy bands, i.e., HR3=(1-2)/(0.3-1).}
\label{fig:hr-orb}
\end{figure*}

\subsubsection{Optical Light Curves and Power Spectra} \label{sec3.12}
The optical light curves of J0503 are shown in the bottom panel of Figure \ref{fig:lc_epic} and Figure \ref{fig:lc_tess} from OM and \textit{TESS}, respectively. The inset of Figure \ref{fig:lc_tess} also shows a close-up of some variability cycles. The long-term variable light curve is shown in Figure \ref{fig:lc_all}, where ASAS-SN and AAVSO data points are plotted together. As explained earlier, we have performed the LS periodogram analysis to search for the periodicity in the data. The LS power spectrum is shown in Figure \ref{fig:ps_all}, where we have marked the positions of all identified frequencies. These frequencies are \orb, \twoorb, \s, and \newf ~and the corresponding periods are given in Table \ref{tab:periods}. Due to the better time-cadence and longer observation duration in \textit{TESS} and AAVSO than available X-ray data, we were able to get more precise values of periods from the optical data. From the present analysis, we found a dominant peak corresponding to the orbital period of 81.65 $\pm$ 0.04 min and 81.63 $\pm$ 0.02 min from \textit{TESS} and AAVSO, respectively, which are similar to that derived from \textit{XMM-Newton} observations. The second dominant peak, which is the second harmonic of the orbital frequency, corresponds to the period (\ptwoo) of 40.82 $\pm$ 0.01 min. In OM data, only \twoorb ~frequency was found to be above the confidence level and the obtained value of \ptwoo is well within a 1$\sigma$ level of the period obtained with the \textit{TESS} and AAVSO. We also refined \ps $\sim$ 65 min available in X-ray data as 65.53 $\pm$ 0.03 min and 65.53 $\pm$ 0.01 min from \textit{TESS} and AAVSO, respectively. In addition to these, a period of 108.22 $\pm$ 0.08 min (\textit{TESS}) and  108.14 $\pm$ 0.03 min (AAVSO) was also found in the power spectrum, which corresponds to the frequency \newf.

% Table-2
\begin{table*}
\begin{center}
\caption{Energy-dependent pulse fractions obtained from spin and orbital phase-folded light curves of J0503 for pole-1 and pole-2.}
\label{tab:pulse-fraction}
\end{center}
\renewcommand{\arraystretch}{1.4}
\begin{tabular}{ccccccc}
\hline
\multirow{2}{*}{Energy Bands} & \multicolumn{6}{c}{Pulse fraction (\%)} \\
\cline{3-7}
\multirow{2}{*}{(keV)} &&  \multicolumn{2}{c}{pole-1} & & \multicolumn{2}{c}{pole-2}\\
\cline{3-4} \cline{6-7}
 && Spin & Orbital && Spin  & Orbital \\
\hline
0.3-1.0 && 53 $\pm$ 8 & 48 $\pm$ 8 && 54 $\pm$ 7 & 48 $\pm$ 6\\
1.0-2.0 && 43 $\pm$ 11 & 43 $\pm$ 10 && 40 $\pm$ 7 & 34 $\pm$ 7\\
2.0-5.0 && 36 $\pm$ 15 & 37 $\pm$ 14 && 32 $\pm$ 7 & 29 $\pm$ 8 \\
5.0-10.0 && --- & --- && 22 $\pm$ 10 & 23 $\pm$ 11 \\
0.3-10.0 && 76 $\pm$ 4 & 70 $\pm$ 4 && 62 $\pm$ 3 & 46 $\pm$ 4\\
\hline
\end{tabular}
\end{table*}

% Figure-11
\begin{figure*}
\centering

\subfigure[]{\includegraphics[width=8cm, height=6cm]{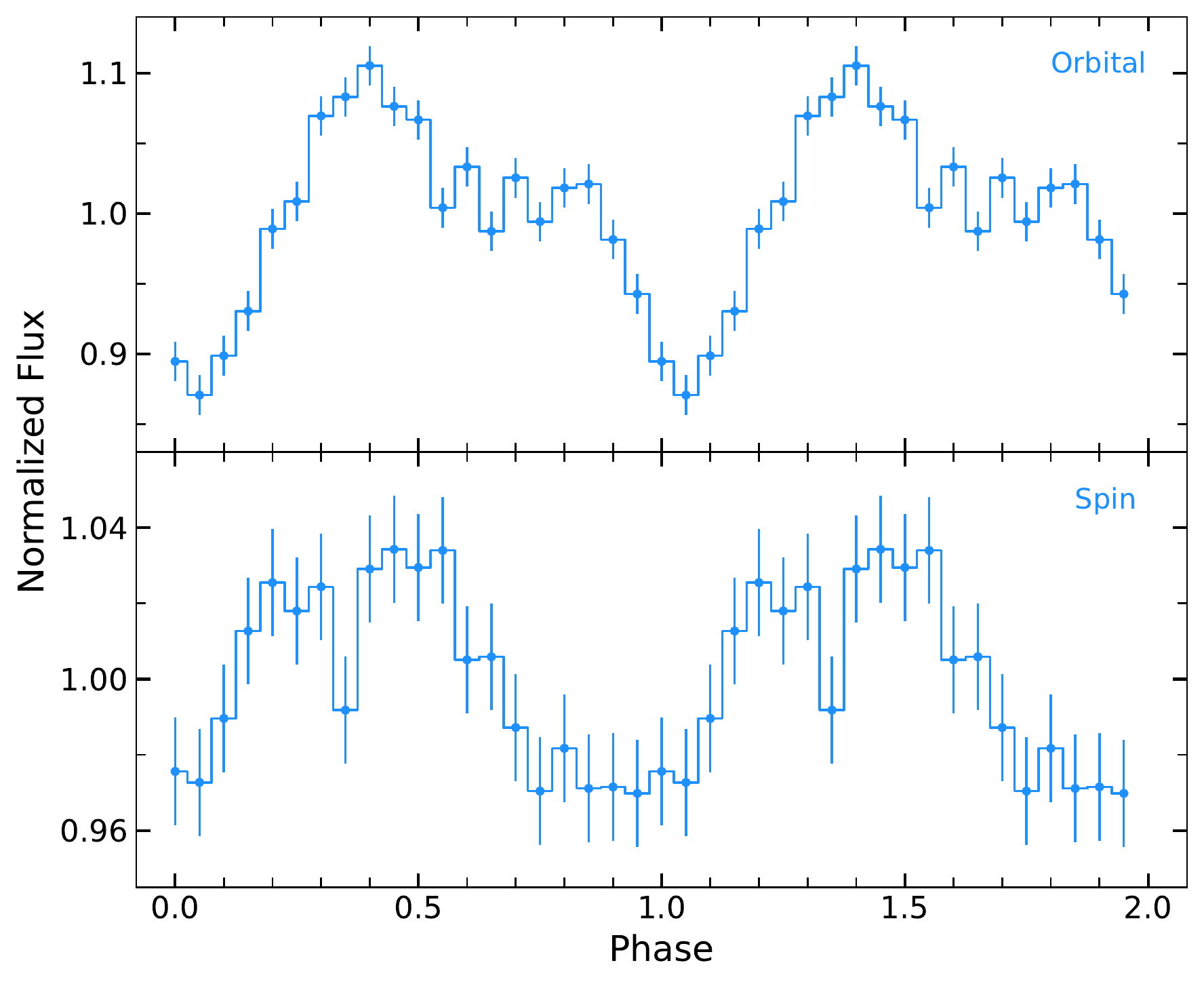}
\label{fig:tess:flc}}
\subfigure[]{\includegraphics[width=8cm, height=6cm]{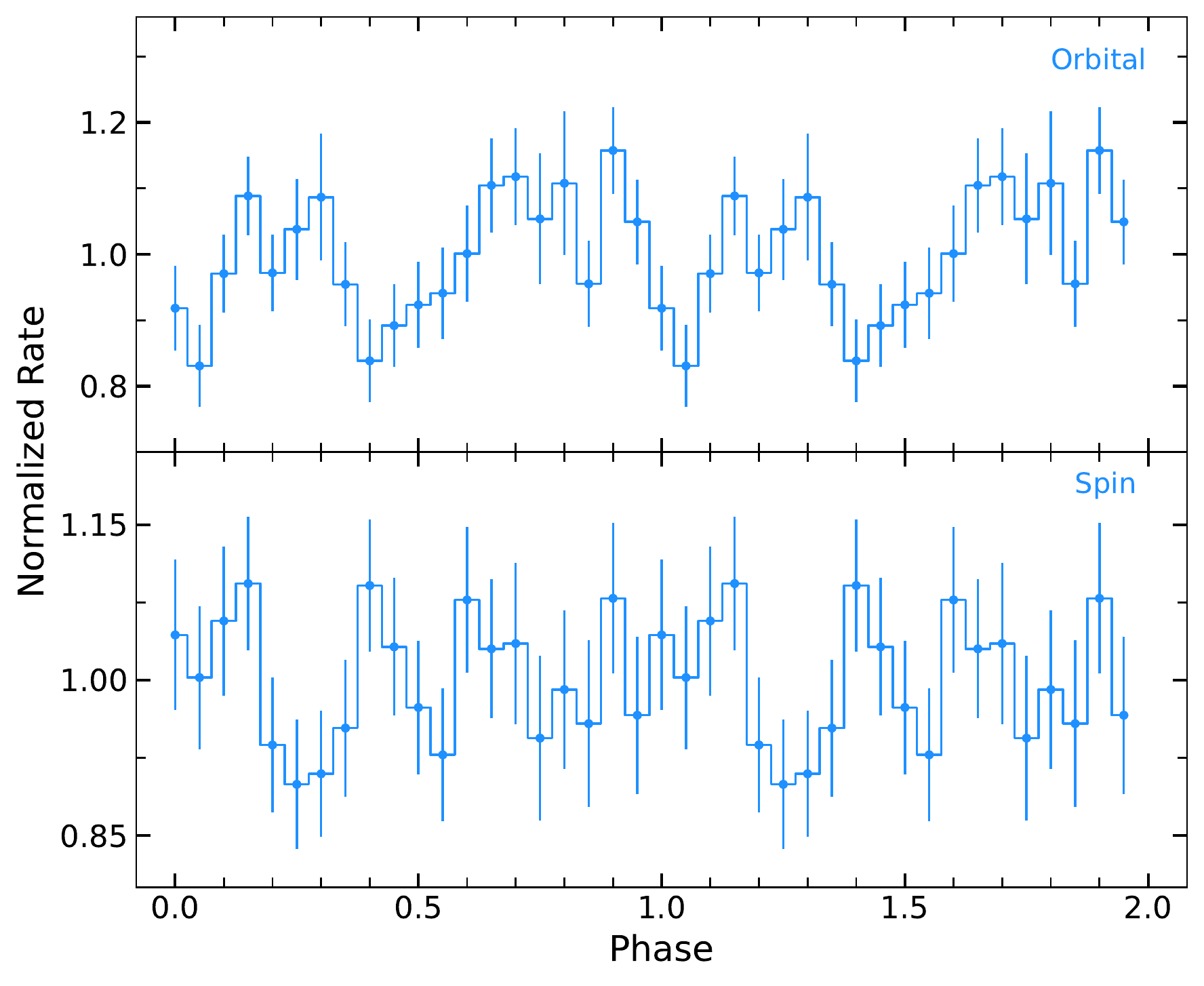}
\label{fig:om:flc}}
\caption{(a) \textit{TESS} and (b) OM folded light curves at orbital and spin periods.  }
\label{fig:opt-flc}
\end{figure*}

\subsubsection{Periodic Intensity Variations}
To explore the periodic variability of J0503, we have folded energy-resolved X-ray and optical light curves over our derived spin and orbital periods. The X-ray and optical light curves were folded with a binning of 25 and 20 points in a phase, respectively.  The reference time for folding was taken to be the first point of \textit{XMM-Newton} observations. The X-ray light curves were folded for pole-1 and pole-2 timings and are shown in Figures \ref{fig:spin-fold} and \ref{fig:orb-fold} for the spin and orbital periods, respectively. The optical spin and orbital folded light curves are represented in Figure \ref{fig:opt-flc}. We have estimated the pulse fraction with $(\rm I_{ max}-I_{min})$/$(\rm I_{max}+I_{min})$ $\times$ 100 \%, where $\rm I_{max}$ and $\rm I_{min}$ are maximum and minimum intensities in a pulse profile, respectively.
The spin pulse profiles for pole-1 and pole-2 timings look  opposite to each other. The pole-1 profiles show broad minima with a single maximum at phase $\sim$ 0.4, whereas the pole-2 profile shows broad maxima and narrow minima at phase $\sim$ 0.88. The derived values of spin pulse fraction in the different energy bands are given in Table \ref{tab:pulse-fraction}. An explicit energy dependency is  seen for both pole-1 and pole-2, where the spin pulsation is more prominent at lower energies. The OM light curve shows no spin modulation; however, in the \textit{TESS} spin folded light curve, the pulse fraction is found to be 3 $\pm$ 1 \%.  We have also examined the hardness ratio (HR) curves for the spin phase defined as HR1, HR2, and HR3 between the hard and soft energy count rates and the corresponding light curves are shown in Figure \ref{fig:hr-spin}. HR1 is the ratio of the count rate in the 5.0-10.0 keV to 2.0-5.0 keV energy band, i.e. HR1=(5-10)/(2-5). HR2 is the ratio of the count rate in the 2.0-5.0 keV to 1.0-2.0 keV energy band, i.e. HR2=(2-5)/(1-2), whereas HR3 is the ratio of the count rate in the 1.0-2.0 keV to 0.3-1.0 keV energy band, i.e. HR3=(1-2)/(0.3-1). All three HR curves for both poles seem to be 180\textdegree ~out of phase to the intensity modulation, i.e. the maximum of the HR curve occurs at the lowest intensity. 

\par Similar to spin folded profiles, the orbital phase folded light curve profiles for pole-1 and pole-2 are found opposite to each other. The pole-1 profiles show broad minima with a single maximum at phase $\sim$ 0.2. On the other hand, pole-2 profiles show broad maxima and narrow minima at phase $\sim$ 0.9. The derived values of orbital pulse fractions in the different energy bands are shown in Table \ref{tab:pulse-fraction}.  A clear energy dependency can be seen for both poles as the orbital pulsations are more prominent at lower energies. Similar to HR curves for the spin phase, we have also determined hardness ratios for the orbital phase. The HR curves for the orbital phase in shown in Figure \ref{fig:hr-orb}. All three HR curves for both poles appear to be opposite to the intensity modulation. Further, OM and \textit{TESS} orbital-folded light curves showed a double-humped profile (see, Figure \ref{fig:opt-flc}). The values of orbital pulse fractions in \textit{TESS} and OM are 12 $\pm$ 1 \% and 16 $\pm$ 4 \%, respectively.

% Figure-12
\begin{figure*}
\centering
\subfigure[]{\includegraphics[width=8.5cm, height=8cm]{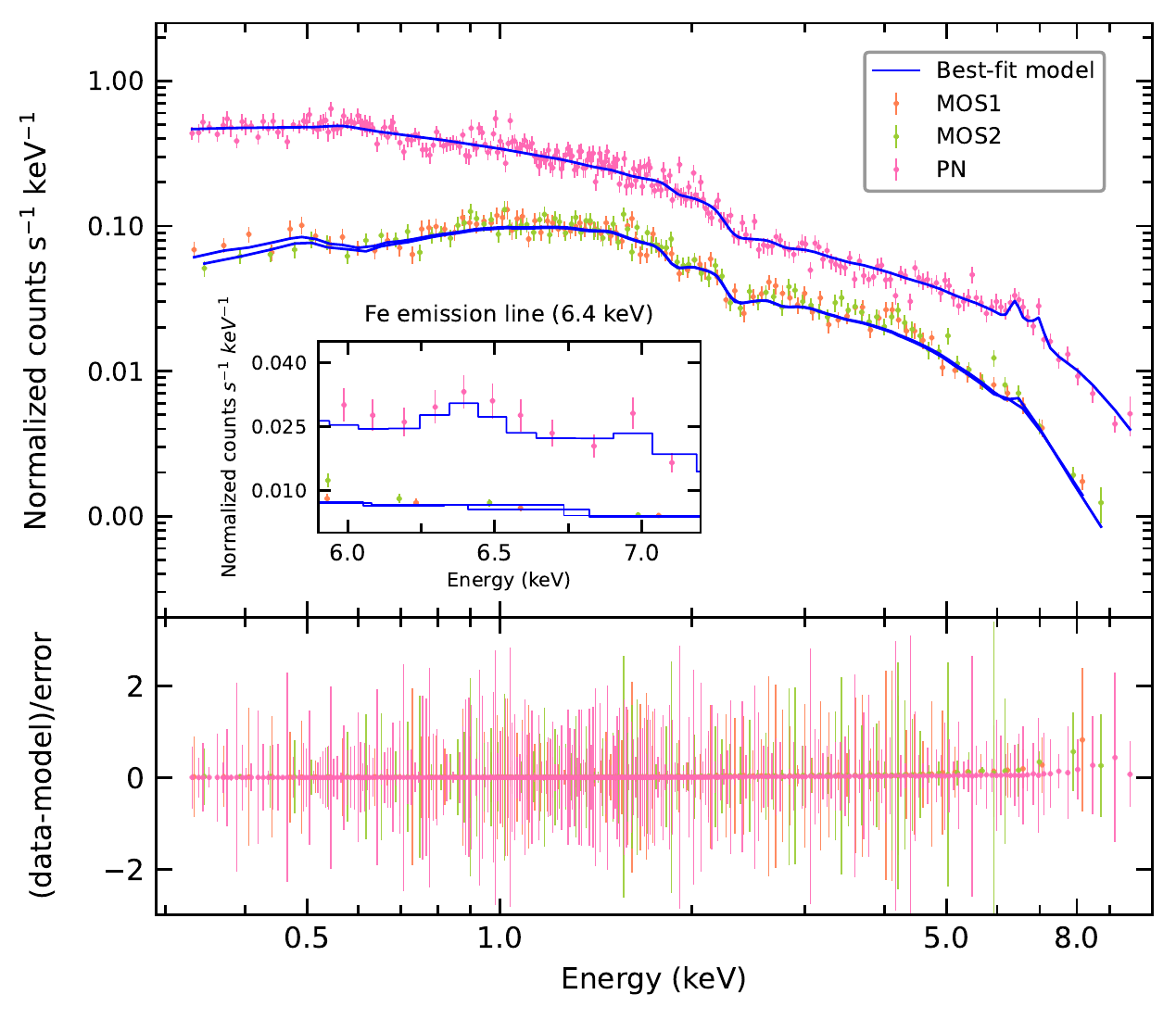}
\label{fig:epic_spec}}
\subfigure[]{\includegraphics[width=8.5cm, height=8cm]{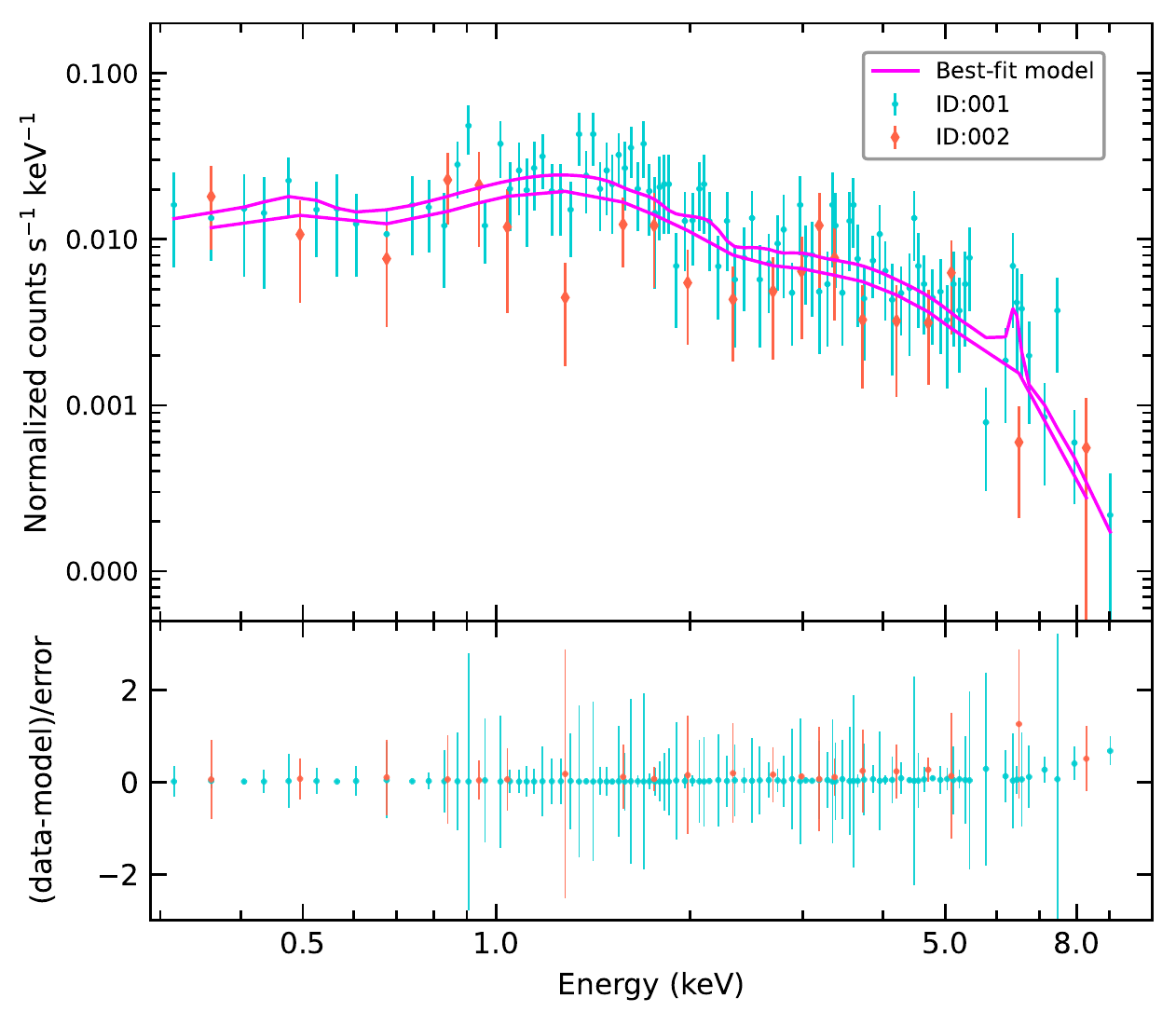}
\label{fig:xrt_spec}}
\caption{X-ray spectra of J0503 as obtained from (a) EPIC (b) XRT along with the best-ﬁt model. The zoomed spectra and the model ﬁt around the Fe-line region are shown in the inset of the ﬁgure. The bottom panel in each Figure shows the $\chi^{2}$ contribution of data points for the best-ﬁtting model in terms of residual.  }
\label{fig:spec}
\end{figure*}

% Table-3
\begin{table*}
\renewcommand{\arraystretch}{1.5}
	\centering
\caption {Spectral parameters as obtained from the best-fit model F= ``phabs$\times$pcfabs$\times$(apec+apec+gauss)'' to the X-ray spectra as obtained from EPIC and XRT (0.3-10.0 keV) observations. All the errors estimated here are with a 90\% confidence range of a single parameter.} \label{tab:avg_spec}
\begin{tabular}{p{0.35\columnwidth}p{0.35\columnwidth}p{0.35\columnwidth}p{0.30\columnwidth}}
\hline
Model & Parameters  &  \textit{XMM-Newton} & \textit{Swift}-XRT \\
\hline
\multirow{2}{*}{pcfabs}& N$_{\rm H,pcf}$ ($10^{22} \rm ~cm^{-2}$)  & $3.8_{-1.1}^{+1.4}$ &   $1.0_{-0.6}^{+3.3}$   \\
& pcf(\%)  & $27_{-6}^{+6} $  & $48_{-21}^{+39} $          \\
\multirow{3}{*}{apec}& T$_{1}$ (eV) &  $153_{-26}^{+20} $  & 153$^{\dagger}$      \\
& Z$_{1}$ (Z$_{\odot}$) &  $0.6_{-0.2}^{+0.3} $  & 0.6$^{\dagger}$  \\
& N$_{1}$ (10$^{-4}$)    &  $1.1_{-0.4}^{+0.9} $  &   $1.9_{-1.9}^{+19.5} $    \\
\multirow{3}{*}{apec} & T$_{2}$ (keV)  & $18.5_{-4.1}^{+6.0} $  & > 11.0      \\
& Z$_{2}$ (Z$_{\odot}$) &  0.6 (tied with Z$_{1}$ )  & 0.6$^{\dagger}$ \\
& N$_{2}$ (10$^{-3}$)   &  $2.2_{-0.1}^{+0.1} $  &   $3.2_{-0.7}^{+1.2} $     \\
gaussian & F$_{\rm g}$ (10$^{-6}$) &    $3.4_{-1.3}^{+1.3} $  &    $0.18_{-0.16}^{+0.23} $    \\
%& EW (eV)   &    $102_{-49}^{+39} $   &    $414_{-21}^{+921} $            \\
\multirow{1}{*}{bolometric flux} & F$_{\rm bol}$ (10$^{-12}$ erg cm$^{-2}$ s$^{-1}$)    &    $8.41_{-0.08}^{+0.08}$  &    $6.1_{-0.5}^{+0.5}$  \\
\multirow{1}{*}{bolometric luminosity} & L$_{\rm bol}$ (10$^{32}$ erg s$^{-1}$)    &    $7.07_{-0.07}^{+0.07}$  &    $5.1_{-0.4}^{+0.4}$  \\
& $\chi_\nu^2$ (dof)                       &  1.08(1115)  & 0.79(336)                  \\
\hline
\end{tabular}

\bigskip
\emph{\textbf{Note.}} $\dagger$ indicates that the parameter value is kept fixed at the value obtained from the EPIC spectral fitting. N$_{\rm H,pcf}$ is the partial covering absorber density i.e., absorption due to the partial covering of the X-ray source by the neutral hydrogen column; pcf is the covering fraction of the partial absorber; T$_{1}$ and T$_{2}$ are the apec temperatures; N$_{1}$ and N$_{2}$ are the normalization constants of apec; Z$_{1}$ is the metal abundance relative to the solar value; F$_{\rm g}$ is the line ﬂux of the Fe K$\rm \alpha$ in terms of photons $\rm cm^{-2} \rm s^{-1}$; F$_{\rm bol}$ is the unabsorbed bolometric ﬂux derived for 0.001–100.0 keV energy band; L$_{\rm bol}$ is the corresponding bolometric luminosity calculated by assuming a distance of 837 pc.
\end{table*}

\subsection{X-ray Spectral Analysis}\label{sec3.2}
The  background-subtracted EPIC PN and MOS spectra of J0503 are shown in Figure \ref{fig:epic_spec}. The X-ray spectral analysis was performed  in the energy range 0.3-10.0 keV using XSPEC version-12.12.0 \citep{1996ASPC..101...17A, 2001ASPC..238..415D}. In order to understand the X-ray emission, we attempted various models or combinations of models to fit the spectra. These models were astrophysical plasma emission code \citep[\texttt{apec};][]{2001ApJ...556L..91S} and cooling-flow plasma emission model \citep[\texttt{mkcflow};][]{1988ASIC..229...53M} along with the \texttt{phabs} component to account for the interstellar absorption. The abundance tables   and the photoelectric absorption cross section `bcmc' were taken from  \citet{2009ARA&A..47..481A} and \citet{1992ApJ...400..699B}, respectively. To account for the emission feature seen near the 6.4 keV in the X-ray spectra (see the inset of Figure \ref{fig:epic_spec}), we have used a \texttt{gaussian} component at the fixed-line energy of 6.4 keV and line width of 0.02 keV along with all the above models. We first employed the model A=\texttt{phabs$\times$(apec+gauss)}, in which the model parameters equivalent hydrogen column density ($\rm N_{H}$) and temperature were pegged at the minimum and the maximum value allowed by the model.
%with a reduced  $\chi^2$ ($\chi_\nu^2$) value of 1.14. 
Therefore, for further spectral fitting, we have fixed the $\rm N_{H}$ value to the total Galactic column in the direction of J0503 of $1.08 \times 10^{20}$ cm$^{-2}$ \citep{2005A&A...440..775K}. With the fixed $\rm N_{H}$ value, we were able to constrain the temperature and other spectral parameters with a reduced $\chi^2$ ($\chi_\nu^2$) value of 1.16.
%Model A with the fixed $\rm N_{H}$ value fitted the spectra with the  $\rm \chi_\nu^2$ value of $\sim$1.16 and we were also able to constrain the temperature and other spectral parameters. 
The majority of the X-ray spectra of MCVs suffer from the local absorbers; therefore, to account for the local absorption effect in the spectral fitting, we have included the model \texttt{pwab} as B=\texttt{phabs$\times$pwab$\times$(apec+gauss)}. The model \texttt{pwab} is a power-law distribution of the covering fraction as a function of maximum equivalent hydrogen column $\rm N_{H,max}$ and the power-law index for the covering fraction $\beta$ \citep{1998MNRAS.298..737D}.  Unfortunately, the spectral fitting could not be constrained with model B. Further, accretion post-shock regions are expected to exhibit a temperature gradient due to the cooling of the gas approaching the WD surface. Therefore, we used the \texttt{mkcflow} component as model C= \texttt{phabs$\times$pwab$\times$(mkcflow+gauss)}, in which we fixed the lower temperature value of \texttt{mkcflow} to the minimum temperature allowed by the model (0.0808 keV). The redshift required in the \texttt{mkcflow} model cannot be zero. It was thus fixed to a value of 1.95 $\times$ 10$^{-7}$ for a Gaia distance of 837$_{-43}^{+60}$ pc \citep{2021AJ....161..147B} and a cosmological Hubble constant of 70 km s$^{-1}$ $\rm Mpc^{-1}$. We used the switch parameter with a value of 2, which determines the spectrum to be computed by using the AtomDB data. With model C, the $\rm N_{H,max}$ value could not be constrained and the higher temperature was pegged at the maximum value allowed by the \texttt{mkcflow} model with a $\chi_\nu^2$ value of 1.31. We, therefore, tried two temperature plasma components as model D= \texttt{phabs$\times$pwab$\times$(apec+apec+gauss)}. Although we faced a similar issue as in model C for $\rm N_{H,max}$, a better $\chi_\nu^2$ value of 1.09 was obtained. Therefore, we replaced the \texttt{pwab} component with partially covering absorption component \texttt{pcfabs} in models C and D. Models E=\texttt{phabs$\times$pcfabs$\times$(mkcflow+gauss)} and F=\texttt{phabs$\times$pcfabs$\times$(apec+apec+gauss)} thus resulted in a significantly better fit than before with slightly improved $\rm \chi_\nu^2$ values of 1.11 and 1.08, respectively. The F-test showed that the model F was more significant than E with an F-statistics of 31.8 and a null hypothesis probability of 2.2 $\times$ 10$^{-8}$. Thus, we adopted the model F as a best-fit model for the spectral fitting of J0503. The unabsorbed bolometric flux in the 0.001-100 keV energy band was also calculated by incorporating the \texttt{cflux} model in the best-fit model F. 

\par The background-subtracted X-ray spectra of J0503 obtained from \textit{Swift-XRT} are shown in Figure \ref{fig:xrt_spec}. The best-fit model F, as discussed earlier, was used for spectral fitting. The low temperature was fixed at a value obtained from the EPIC fitting. In this way, we have obtained a lower limit of high temperature for XRT spectra. The spectral parameters derived from the simultaneous ﬁtting to both EPIC PN and MOS spectra and XRT spectra of both observations using the best-fit model F together with the 90\% confidence limit for a single parameter are given in Table \ref{tab:avg_spec}.

\section{Discussion}\label{sec4}

We have carried out X-ray and optical timing analyses and X-ray spectral analysis of a probable IP J0503. From the long-term \textit{TESS} observations, we refined the \po of the system as 81.65 $\pm$ 0.04 min. We speculate that the 65.53 $\pm$ 0.03 min periodicity is associated with the \ps of the WD. Considering the previously mentioned values of \po and \ps, the longest duration periodicity present in the X-ray power spectra could be associated with the \pb. The presence of these periodicities in the X-ray and optical data indicates the IP nature of J0503. Adopting these values of \po and \ps, J0503 might fall in the category of nearly synchronous IPs with Paloma \citep{2007A&A...473..511S, 2016ApJ...830...56J} with \ps/\po = 0.802. If it is truly a nearly synchronous IP, it is an important addition to this class because J0503 is the first one with an orbital period less than the `period gap'. There is another category of IPs known as EX Hya-like systems, for which \ps/\po > 0.1 and \po < 2 h with systems EX Hya \citep{1980A&A....85..106V} and V1025 Cen \citep{1998MNRAS.299...83B}. For EX Hya and V1025 Cen, \ps/\po is 0.68 and 0.42, respectively, which is less than the value $\sim$ 0.8 obtained for J0503. Moreover, the evolution of IPs can be understood from the distribution of their spin and orbital periods in the \ps - \po ~plane and the degree of synchronisation with the orbital period. The asynchronicity parameter (1-\ps/\po) for J0503 was found to be $\sim$ 19.8 \%, which suggests that it is slightly away from the line of synchronisation. The obtained value of \ps/\po = 0.802 also satisfies the synchronisation condition \ps/\po > 0.6 derived by \citet{2004ApJ...614..349N}. The reason for J0503 not being synchronised could be that the secondary star has a low magnetic moment and therefore, it is unable to come into synchronism, similar to the nearly synchronous IPs and EX Hya-like systems as suggested by \citet{2004ApJ...614..349N}.

\subsection{Governing accretion mechanism}
The X-ray and optical power spectra of IPs serve as an important diagnostic to understand the mode of accretion in these systems. The X-ray power spectra of J0503 have shown the presence of major frequencies such as \s, \orb, \be, \twobe, \threebe, \beplus, and \sone, while the optical power spectra have shown only \orb, \twoorb, \s, and \newf ~frequencies. These results somewhat match model 1 of \citet{2022ApJ...934..123H} in terms of the identification of \s frequency. Still, the difference lies in the absence of \be, its harmonics, and \sone in the X-ray power spectra of the author, which are the primary frequency components to look for while explaining accretion scenarios in IPs. The possible mechanisms of the presence of these frequencies are described in the forthcoming paragraphs.
\par The orbital modulation in the X-ray light curves of IPs is generally explained by any of the following mechanisms: (i) Whenever the accretion stream impacts with the disc or the magnetosphere, depending on the mode of accretion, it throws material out of the orbital plane. Obscuration of X-rays by such material rotating in the binary frame might produce orbital modulation. (ii) photoelectric absorption of X-rays by the material rotating in the binary frame also produces the orbital modulation, which is energy-dependent. (iii) In the disc-overflow accretion scenario, an interaction between intrinsic modulations at the spin and beat periods leads to the apparent modulation at the orbital period. As we will proceed, we will see that the matter is not accreting via disc-overflow accretion. Further, the energy-dependent orbital phase folded light curves for pole-1 and pole-2 show dominant modulation in softer energy bands and modulation decreases with increasing energy. Therefore, the most likely mechanism for the orbital modulation in X-rays for J0503 is the photoelectric absorption of X-rays by the material rotating in the binary frame. Whereas, the broad minima in the folded light curves of pole-1 also suggest that the orbital modulation could be due to the obscuration by the material in the binary frame along with photoelectric absorption. 

\par On the other hand, X-ray modulation at the spin frequency is the definitive characteristic of the IPs, which can arise due to the two mechanisms: (i) photoelectric absorption and electron scattering in the infalling material  and (ii) self-occultation of emission regions by the WD. The values of spin pulse fraction for pole-1 and pole-2 are energy-dependent, which decrease with increasing energy. Therefore photoelectric absorption in the accretion flow could be the main reason for the spin modulation in J0503, similar to the majority of IPs validating the ``accretion curtain'' model \citep[see][for details]{1989MNRAS.237..853N}. Further, optical spin pulsation is seen due to the absorption of X-rays by a structure locked to the WD, such as an accretion curtain \citep{1986MNRAS.219..347W}.

\par We do not see equal power at \be and \beplus; therefore, we can safely assume that the beat period is not entirely caused by the amplitude modulation of spin frequency at the orbital period. A slight asymmetry in the powers of \be and \beplus frequencies in the 0.3-10.0 keV band suggest that  an intrinsic beat modulation is present in the system due to the accretion taking place via a stream. The modulation at the \be frequency arises as the accretion stream flips between the magnetic poles twice for every rotation of the WD with respect to the binary frame.   

\par To explore the possibility of dominant accretion mechanism, we adopt the models given by \citet{1992MNRAS.255...83W} and \citet{1999MNRAS.309..517F}. Considering the symmetric model given by \citet{1992MNRAS.255...83W} and adopting the same terminology used, with high values of the angle of inclination (i) and co-latitude (m), for which the condition $\rm i +\, \rm m$ > 90\textdegree $+\, \beta$ satisfies, stream-fed accretion produces modulation at frequencies \be, \orb, \sone, \stwo ~etc. The strongest of these would be \s $\pm$ \be = \sone or \orb. While the asymmetric model produces similar power spectra with an addition of the presence of \s peak.  The model includes asymmetry by differing in pole cap luminosities and size. Further, the power of the spin peak is proportional to the degree of asymmetry introduced. Extreme differences between the pole caps can cause \s to become dominant and modulation at the weaker \beplus  sideband also becomes evident. We emphasize that the presence of \sone in the power spectra of J0503 can not be considered associated with the orbital modulation of 2$\omega$ because orbital modulation of a weak signal at 2$\omega$ can not produce a much higher signal at \sone. According to \citet{1992MNRAS.255...83W}, the best way to distinguish between the stream-fed and disc-fed systems is in the hard X-ray regimes. They have shown that the stream-fed systems produce power at \s, \be, and \sone although one of the latter two should always be present, while disc-accreting systems produce power at \s only. We have seen that in hard X-ray bands (> 5.0 keV), only \be, \orb, and \beplus are present in the power spectra of J0503, which indicate that the stream-fed accretion is the feasible accretion scenario for J0503. Moreover, \citet{1999MNRAS.309..517F} have shown the importance of stream extensions while determining the characteristics of power spectra. As the azimuthal extension of the source of matter in the orbital plane increases, \s becomes prominent even in the stream-fed accretions and \be could be absent in this scenario \citep[see the top panel of figure 7 for $\delta$$\phi$=180\textdegree ~in][]{1999MNRAS.309..517F}, which we have also seen in the optical power spectra of J0503. All these features can be attributed that the J0503 might be predominantly accreting via stream during present observations.

\subsection{The post-shock region: two-temperature structure}
The average EPIC spectra suggest that the X-ray post-shock emitting region has a two-temperature structure which is characterized by a low ($\sim$ 150 eV) and a high ($\sim$ 18.5 keV) component with a partial covering absorber of equivalent hydrogen column of $\sim$ 3.8 $\times$ 10$^{22}$ cm$^{-2}$ and a covering fraction of $\sim$ 27 \%. Assuming the maximum temperature (18.5 keV) derived from the spectral fitting as the shock temperature and adopting the WD mass-radius relationship of \citet{1972ApJ...175..417N}, we have calculated WD mass to be 0.54$^{+0.18}_{-0.12}$ M$\odot$. Similar low mass values have been found in EX Hya \citep{1997ApJ...474..774F} and HT Cam \citep{2005A&A...437..935D}. Using the value of unabsorbed bolometric luminosity (7.07 $\times$ 10$^{32}$ erg s$^{-1}$) derived from the spectral fitting and equating this to the accretion luminosity L$_{\rm acc}$ = GM$_{\rm WD}$ $\dot{\rm M}$/R$_{\rm WD}$, we have calculated mass accretion rate ($\dot{\rm M}$) to be 1.44 $\times$ 10$^{-10}$ M$\odot$ yr$^{-1}$,
%(0.9135 $\times$ 10$^{16}$ g s$^{-1}$),
which is in the range of the expected mass accretion rate for IPs. The shock height in an accretion column can be described by the relationship H = 5.45 $\times$ 10$^{8}$ $\dot{\rm M}_{16}^{-1}$ f$_{-2}$ M$_{\rm WD}^{3/2}$ R$_{\rm WD}^{1/2}$ \citep{1992apa..book.....F}, where $\dot{\rm M}_{16}^{-1}$ is the mass accretion rate in units of 10$^{16}$ g s$^{-1}$, f$_{-2}$ is the fractional area in units of 10$^{-2}$, M$_{\rm WD}$ and R$_{\rm WD}$ are the mass and the radius of white dwarf in solar units, respectively. We have assumed the fractional area to be 10$^{-3}$ \citep[see][for details]{1992MNRAS.254..493R}. Using the derived values of $\dot{\rm M}$, M$_{\rm WD}$, and R$_{\rm WD}$ the value of shock height was thus found to be 2.76 $\times$ 10$^{6}$ cm or 0.003 R$_{\rm WD}$. 

\section{Conclusions}\label{sec5}
We conclude our findings as:
\begin{itemize}
    \item[1.] The orbital period derived from X-ray, \textit{TESS}, and AAVSO data is well consistent with the previously reported value.
    \item[2.] The $\sim$ 65 min periodicity found in X-ray and optical data could be attributed to the spin period of the WD. If it is so, then J0503 would be the first nearly synchronous IP (\ps/\po $\sim$ 0.8) below the period gap.
    \item[3.] The presence of \s, \orb, \be, \twobe, \threebe, \beplus, and \sone ~frequencies indicates that the J0503 might be accreting predominantly via stream-fed accretion.
    \item[4.] The energy-dependent spin pulsations for pole-1 and pole-2 are due to the photoelectric absorption in the accretion flow, which is one of the unique characteristics of the majority of the IPs.
    \item[5.] The post-shock emitting region is well explained by the presence of two temperatures: low ($\sim$ 150 eV) and high ($\sim$ 18.5 keV).
    \item[6.] A partial covering absorber of equivalent hydrogen column of $\sim$ 3.8 $\times$ 10$^{22}$ cm$^{-2}$ and a covering fraction of $\sim$ 27 \% was found to be reasonable to explain the X-ray spectrum.
    \item[7.] We have estimated WD mass, mass accretion rate, and shock height to be $\sim$ 0.54 M$\odot$, 1.44 $\times$ 10$^{-10}$ M$\odot$ yr$^{-1}$, and 2.76 $\times$ 10$^{6}$ cm (0.003 R$_{\rm WD}$), respectively.
\end{itemize}

\section{Acknowledgements}
We thank the anonymous referee for providing useful comments and
suggestions that led to the significant improvement of the quality of the paper. This research has made use of the data obtained with \textit{XMM-Newton}, an ESA science mission with instruments and contributions directly funded by ESA Member States and NASA. This paper also includes data collected with the \textit{TESS} mission, obtained from the MAST data archive at the Space Telescope Science Institute (STScI). Funding for the \textit{TESS} mission is provided by the NASA Explorer Program. We acknowledge with thanks the variable star observations from the AAVSO International Database contributed by observers worldwide and used in this research. This research has made use of the XRT Data Analysis Software (XRTDAS) developed under the responsibility of the ASI Science Data Center (ASDC), Italy.

\section{DATA AVAILABILITY}
The \textit{XMM-Newton} and \textit{Swift} data used for analysis in this article are publicly available in NASA’s High Energy Astrophysics Science Archive Research Center (HEASARC) archive (\url{https://heasarc.gsfc.nasa.gov/docs/archive.html}). The \textit{TESS} data sets are publicly available in the \textit{TESS} data archive at \url{https://archive.stsci.edu/missions-and-data/tess}. The AAVSO and ASAS-SN data sets are available at \url{https://www.aavso.org/data-download} and  \url{https://asas-sn.osu.edu/variables}, respectively.

\bibliographystyle{mnras}
\bibliography{ref.bib}

\appendix

\label{lastpage}
\end{document}